PSL
UNIVERSITÉ PARIS

**HABILITATION À DIRIGER DES RECHERCHES**

**DE L'UNIVERSITÉ PSL**

Présentée à l'institut Langevin

# Doppler holography for ophthalmology

# Holographie Doppler pour l'ophtalmologie

Présentation des travaux par

**Michael ATLAN**

Le 21 septembre 2023

Discipline

**Optique**

Composition du jury :

| | |
|---|---|
| Olivier HAEBERLE<br>Professeur, Université de Haute-Alsace | *Rapporteur* |
| Corinne FOURNIER<br>Maître de conférences, Université Jean Monnet | *Rapporteur* |
| Laurent MUGNIER<br>Directeur de recherche, ONERA | *Rapporteur* |
| Michel GROSS<br>Directeur de recherche, CNRS | *Examinateur* |
| Alain GAUDRIC<br>Professeur, Université Paris Cité | *Examinateur* |
| Michel PAQUES<br>Professeur, Sorbonne Université | *Examinateur* |
| Ramin TADAYONI<br>Professeur, Université Paris Cité | *Examinateur* |
| Denis LEBRUN<br>Professeur, Université de Rouen | *Examinateur* |
| Pascal PICART<br>Professeur, Université du Mans | *Examinateur* |

Institut Langevin
ONDES ET IMAGES

# Doppler holography for ophthalmology

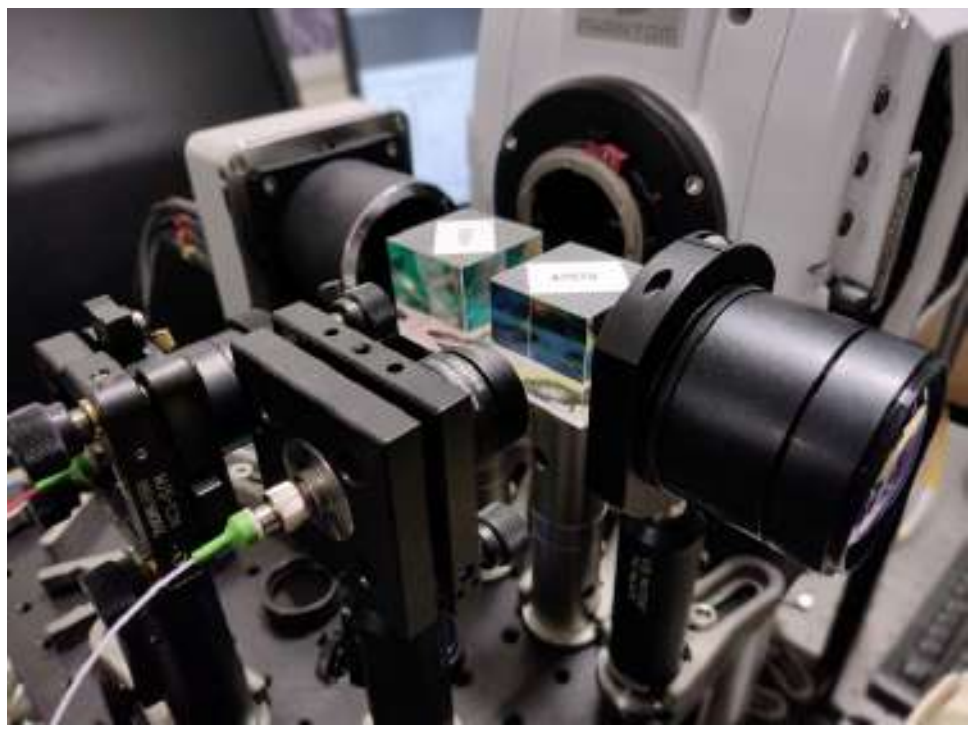
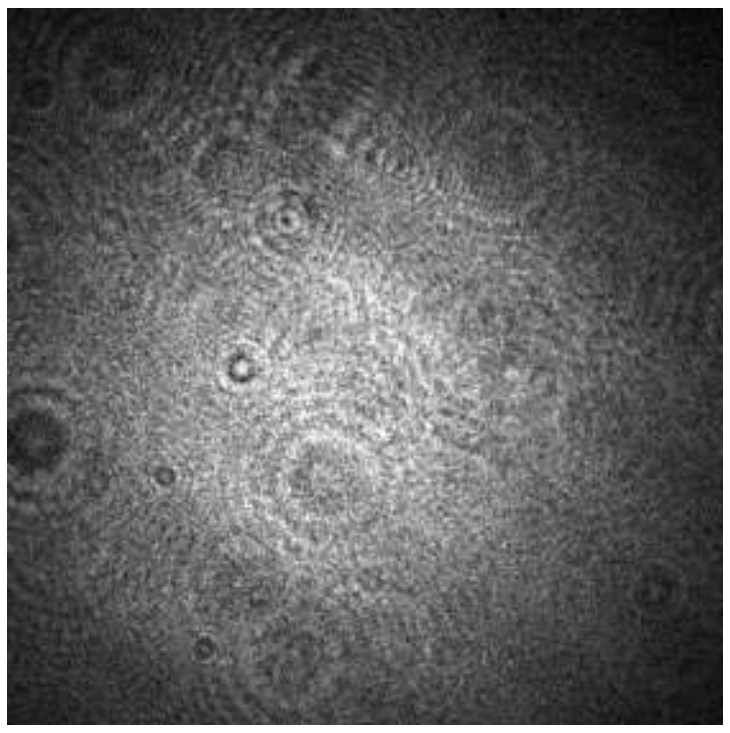
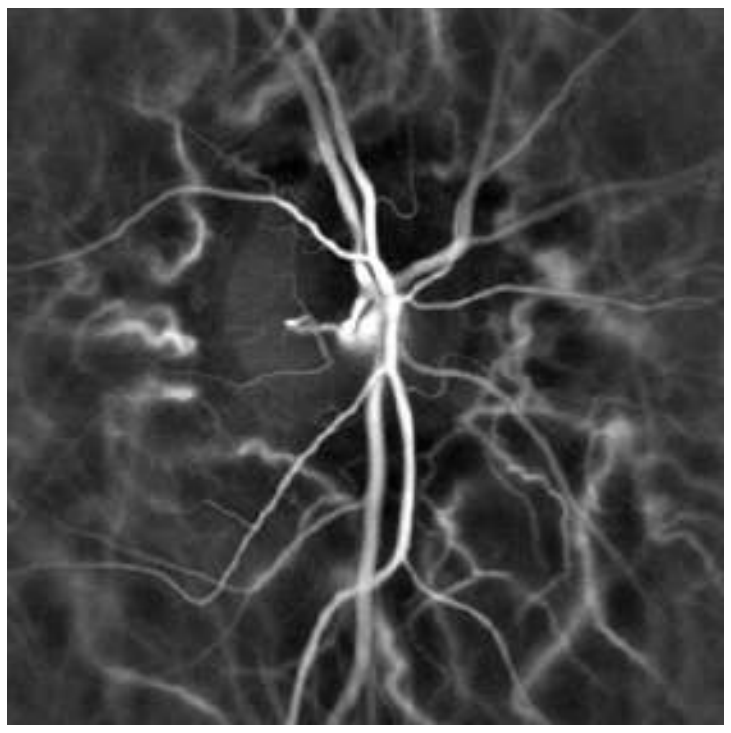

Michael Atlan

ESPCI PARIS | PSL

# Table of contents

# Abstract

A comprehensive assessment of retinal health demands reliable and precise methods to measure localized blood perfusion. Despite considerable advancements in imaging techniques, such as indocyanine green and fluorescein angiography, along with optical coherence tomography angiography, their capacity to monitor blood flow dynamics across the cardiac cycle faces significant limitations. For more effective care of those with ocular conditions, innovating new approaches is paramount. Doppler holography, an emerging non-invasive optical imaging technique, meets this challenge by offering high temporal resolution imaging of retinal and choroidal blood flow. Now a burgeoning interdisciplinary research field, Doppler holography intertwines functional optical imaging system design, high-performance computing, and clinical investigation. Through collaborative efforts among universities, industry partners, and ophthalmic clinics, a network for its advancement and application is taking shape. This endeavor promises to propel the discovery of novel functional biomarkers, transforming the diagnosis and treatment of retinal diseases, refining disease severity categorization, and enhancing therapeutic monitoring—ultimately leading to improved healthcare outcomes.

Une évaluation complète de la santé rétinienne nécessite des méthodes fiables et précises pour mesurer la perfusion sanguine localisée. Malgré des avancées considérables dans les techniques d'imagerie telles que l'angiographie au vert d'indocyanine et à la fluorescéine, ainsi que l'angiographie par tomographie en cohérence optique, leur capacité à surveiller la dynamique du flux sanguin tout au long du cycle cardiaque présente d'importantes limitations. Pour une prise en charge plus efficace des personnes souffrant de conditions oculaires, innover en proposant de nouvelles approches est primordial. L'holographie Doppler, une technique d'imagerie optique non invasive émergente, relève ce défi en offrant une imagerie à haute résolution temporelle du flux sanguin rétinien et choroïdien. Désormais un domaine de recherche interdisciplinaire en plein essor, l'holographie Doppler entrelace la conception de systèmes d'imagerie optique fonctionnelle, l'informatique haute performance et l'investigation clinique. Grâce aux efforts collaboratifs entre universités, partenaires industriels et cliniques ophtalmologiques, un réseau pour son développement et son application se forme. Cette initiative promet de stimuler la découverte de nouveaux biomarqueurs fonctionnels, transformant le diagnostic et le traitement des maladies rétiniennes, affinant la catégorisation de la gravité des maladies, et améliorant la surveillance thérapeutique - conduisant finalement à de meilleurs résultats de soins de santé.

# Introduction

Ophthalmic conditions are an escalating global health concern. By 2050, visual impairments and blindness are anticipated to impact an astounding 100 million and 600 million individuals respectively. A significant proportion of these conditions, such as glaucoma, macular degeneration, and diabetic retinopathy, are directly associated with changes in blood flow within retinal microvascular networks. While imaging techniques like indocyanine green (ICG), fluorescein angiography, and optical coherence tomography (OCT) angiography have evolved, many still grapple with accurately capturing blood flow dynamics. The pressing demand now is for non-invasive tools capable of precisely evaluating essential quantitative biomarkers—like blood flow velocity, blood volume rate, arterial wall stiffness, arterial resistivity index, and blood viscosity within retinal vessels. Such innovations hold the promise to redefine ocular disease management, significantly improving patient outcomes.

The creation of high-speed digital holographic imaging devices for ophthalmology is pivotal in addressing the demand for non-invasive diagnosis and monitoring of ocular conditions. Doppler holography stands out, especially concerning blood flow imaging. The fruits of these technological advancements can be seen in instruments developed for clinical research at both the Quinze-Vingts Hospital and the Rothschild Foundation Hospital in Paris in recent years. Leveraging these state-of-the-art technologies, we can now detect alterations in retinal vascular networks earlier and consistently monitor ocular pathology progression and treatment. These instruments offer fresh avenues to enhance the management and prevention of complications for those suffering from ocular diseases. Notably, we have successfully pioneered ultra-high data rate clinical holographic imaging devices. These robust devices are in the process of being transferred to industry partners, as well as to numerous global clinics and academic laboratories. The University of Pittsburgh Medical Center (UPMC) in the USA notably replicated the first functional instrument in 2022, signaling a landmark moment in the computational imaging of optical imagery in ophthalmology.

During its inception, near-infrared Doppler holography paired with ultrahigh-speed cameras faced formidable technological barriers, particularly concerning eye fundus imaging. These hurdles included managing low-frequency spurious interferometric signal contributions, ensuring eye safety compliance, and attaining high bitrate hologram real-time rendering. Overcoming these, singular value decomposition filtering was employed to extract superior backscattered optical wave signals using on-axis interferometry, thus facilitating precise Doppler imaging of the retina. An innovative diffuse laser illumination technique was also introduced, ensuring alignment with European ISO 15004-2-2007 and American ANSI Z136.1-2014 safety standards. A significant milestone was the integration of a high-speed streaming camera, achieving real-time ultra-fast medical holographic Doppler imaging—a world-first accomplishment.

Doppler holography in ophthalmology emerges as an interdisciplinary research arena, melding the design of functional optical imaging systems and high-performance computing with clinical research, further bolstered by collaboratively developed open-source software. Collaborative efforts across universities, industrial partners, and ophthalmic clinics are forging a network for computational holographic ophthalmic imaging. This endeavor has the potential to introduce less invasive, robust, and adaptable ultra-high-speed ophthalmic imaging devices that showcase functional contrasts supplementing state-of-the-art dye-based and OCT angiography methods. The research trajectory in Doppler holography is now pivoting towards uncovering new functional biomarkers to refine non-invasive pathology severity classification and optimize therapeutic monitoring, ultimately aiming to elevate healthcare outcomes.

# Accompanying note

## Curriculum vitæ

### Civil State

Last Name: Atlan

First Name: Michael

Date of Birth: 2nd January 1978

Nationality: French

ORCID Number: 0000-0003-1899-6419

Website: Langevin Institute

Website: Paris Eye Imaging

Source code repository : digital holography

## Resume

I am a researcher with a focus on experimental optical physics, instrumentation, and computational imaging at the French National Center for Scientific Research (CNRS). I hold positions at the Langevin Institute (CNRS UMR 7587), the Quinze-Vingts National Eye Hospital (INSERM CIC 1423), and the Fondation Adolphe de Rothschild Hospital in Paris. My expertise lies in the development of coherent-light devices and advanced numerical processing methods for laser Doppler vibrometry, microscopy, and low-light blood flow imaging. In ophthalmology, I am particularly interested in the development of methodologies that involve optically-acquired digital holograms for sensitive, high throughput coherent detection, and their use for computed imaging of both the anterior and posterior segments of the eye. I serve as the project manager for holovibes, a versatile real-time digital hologram rendering software, and I lead the development of holowaves, a computational image rendering prototyping platform. Beyond my research contributions, I am a co-founder and president of the digital holography foundation. My dedication to the advancement of the field is reflected in my list of peer-reviewed journal articles, patents, and book chapters on the subject of digital holography.

## Education and research experience

- 2019- : Part time work at Rothschild Foundation hospital. 29 rue Manin - 75019 Paris.
- 2016- : Part time work at Clinical Investigation Center of the Hospital "Hôpital des 15-20." CIC 1423. CNRS UMR 7210 Vision Institute, INSERM UMR-S 968, UPMC. 75654 Paris Cedex 13.
- October 2013: CR1 CNRS (Researcher) at Langevin Institute - ESPCI – CNRS UMR 7587.
- October 2009: CR2 CNRS (Researcher) at Langevin Institute - ESPCI – CNRS UMR 7587.
- October 2008: CR2 CNRS Intern at ESPCI, LPEM CNRS UPR 5.
- February 2008 – September 2008: CNRS Postdoctoral Researcher at IJM UMR 7592, Department of Cell Biology.
- February 2007 – January 2008: CNRS Postdoctoral Researcher at ENS, Kastler Brossel Laboratory UMR 8552.
- November 2006 - January 2007: CNRS Fixed-term Contract at ESPCI, Optics Laboratory UPR 5.
- September 2006: Visiting Researcher at NICHD, NIH, Washington (USA).
- February 2006 - September 2006: Postdoctoral Researcher at the University of Texas Austin (USA).
- November 2005 - January 2006: CNRS Research Engineer Fixed-term Contract at ESPCI, Optics Laboratory UPR5.
- October 2002 - September 2005: Ph.D. at the Laboratory of Physical Optics, ESPCI, Paris VI University.

# Publications

# Patents

1. M. Atlan, M. Gross. "Holography device." CNRS Patent. Filed on June 19, 2009. Registration number: FR2947064B1.
2. M. Atlan, M. Gross. "Holography device." CNRS/ENS Patent. Filed on July 23, 2010. Application number: FR2963117A1. Patent abandoned by CNRS.

3. M. Atlan, B. Samson. "Off-axis digital holography." CNRS Patent. Filed on November 15, 2012. Application number: FR1260883. PCT: WO2014076251A1. Europe: EP2920652A1. USA: US9733064B2.
4. M. Atlan, J.P. Huignard. "Method and device for holographic microscopy with Bragg gratings." Patent application filed on March 31, 2017. INPI application number: 17 52775. PCT application number: WO2018178366Al. Europe: EP2915009B1. USA: US10120334B2.
5. Léo Puyo, Michael Atlan. "Methods and devices for full-field ocular blood flow imaging." U.S. Patent 11,457,806, issued on October 4, 2022.
6. G. Laloy-Borgna, L. Puyo, S. Catheline, M. Atlan. National Institute of Health and Medical Research (INSERM) PCT/EP2022/064318, filed on May 25, 2022, published on December 1, 2022.

# Software

I have initiated the creation of the digital holography open source foundation to support the development and sharing of computational imaging software. Its primary mission is to ensure the development and reliability of open-source software for digital hologram rendering and analysis. The foundation's supported applications encompass a broad range of use cases and partners, including clinical-grade real-time, high bitrate Doppler holography and various other computational imaging modalities. The objective of this initiative is to pool and disseminate accumulated expertise in the realm of high-speed digital holography. By doing so, we aim to bolster the research and development of open-source imaging software, while simultaneously providing education and training opportunities for student programmers in parallel computing. The open-source approach contributes to promoting the accessibility of advanced, reliable holographic imaging software tools for researchers, developers, and medical professionals worldwide.

Details about the foundation:

- Name: Digital Holography Non-profit Foundation

- Date of Registration: September 2021

- Registration Numbers: RNA W751262092; SIRET 90325763200013

- Website: www.digitalholography.org

- GitHub: https://github.com/DigitalHolography

# Funding

Major milestones and funding for Doppler Holography. Unless stated, I am the primary investigator (PI).

- 2003: First Doppler holographic measurement, ESPCI-ENS LKB [45].
- 2009: ESPCI Doppler holography internal grant. €34,000.
- 2009: CNANO IDF BioRhéoPassive. €40,000.
- 2009: ANR AngioDoppler (2009-2012). €237,000.
- 2010: First Doppler holographic measurement of the rodent retina [28].
- 2010: Pierre-Gilles de Gennes Foundation (FPGG). €120,000.
- 2011: AIMA Lab IDF grant. €30,000.
- 2011-2013: ANR prematuration VideoLaserDoppler. €347,000.
- 2014-2020: ERC Synergy HELMHOLTZ. ESPCI-1520. PIs: Mr. Fink & Mr. J.-A. Sahel. WP Doppler holography ~€500,000.
- 2017: First Doppler hologram of the human retina [9].
- 2018: IHU FOReSIGHT ANR-18-IAHU-0001 (15-20) i2Eye. PI: Michel Paques WP3 Doppler holography €235,000.
- 2019: SESAME IDF Region: 4DEye-EX047007. PI: Michel Paques. WP Doppler holography. ~€250,000.

- 2020: First real time holography device for human retina at 2,000 frames per second.
- 2021: ½ Thesis funding: Labex WIFI Langevin Institute (+ ERC OptoRetina PI : Kate Grieve).
- 2021: LABEX WIFI AAP UltraFastHolography. €46,000.
- 2021-2023: PSL Qlife UltrafastHolography project Qlife-ESPCI-02-2021. €90,000.
- 2022: First real time holography device for human retina at 20,000 frames per second.
- 2022: ANR LIDARO. ESPCI, QM, 15-20, FOR. €717,000.
- 2023: QM, ESPCI, 15-20, FOR. BPI HoloDoppler. PI: David Pureur. €2,560,000. Request in progress.

# Supervision and co-supervision of PhD students

- M. Léo Puyo. "Clinical application of laser Doppler holography in ophthalmology" (thesis started in October 2016, defended in September 2019). Director : Mathias Fink
- Ms. Julie Rivet. "Non-iterative methods for image improvement in digital holography of the retina" (thesis started in August 2017, defended in July 2020). Director : Thierry Géraud.
- Ms. Zosia Bratasz. Title to be determined (thesis started in September 2021). Director : Kate Grieve.
- Ms. Julia Sverdlin. Title to be determined (thesis started in January 2022). CIFRE contract with EssilorLuxottica. Director : Vincent Borderie.
- M. Yohan Blazy. Title to be determined (thesis started in October 2022). CIFRE contract with Lumibird Quantel Medical.

- M. Olivier Martinache. Title to be determined (thesis started in October 2023). Director : Pedro Baraçal de Mecê.

# Teaching experience

- 2010 - Present: Supervision of an average of 6 trainees per year. Spring/Summer: Instrumentation, measurements, medical imaging, and algorithm design for analysis. Autumn/Winter: C++/CUDA programming for high throughput, real time parallel computing.

- 2003 - 2009: Part-time lecturer. Development and supervision of experimental projects (TP). Topics included "Classical and New Microscopies," "Digital Holography," and "Classical Holography." (360 hours of teaching at BAC+3/4 level). École Centrale Paris, Châtenay-Malabry.

- 2006: Qualification for the position of Associate Professor in the field of Diluted Media and Optics (Section 30). Qualification number: 06230165322.

# Organization of conferences

- Sixth edition of the French-speaking Congress on Digital Holography, Holophi6, in Paris on November 30th and December 1st, 2022. This congress aimed to bring together and foster connections within the French scientific community dedicated to the development and applications of digital holography.

# Clinical research

## Contributions to clinical trial protocol design and ocular safety norms compliance

- IMAMODE clinical trial. Registration number: Clinical trials: NCT04129021 registered on 16/10/2019. PI: Michel Paques. Registration number: FR: ID-RCB:

2019-A00942-55 registered on 01/04/2019. CPP Sud-Est III: 2019-021B. ANSM No. IRDCB: 2019-A00942-55.

- IREDO Protocol for clinical investigation on a medical device. (Case 4.4: IC on CE-unmarked MD without CE marking objective or establishment of conformity) . PI: Sophie Bonnin. Establishment of an eye image database using laser Doppler holography. Research code: SBN_2022_9. IDRCB number: 2022-A01593-40. Application under review.
- STUDY22100132. Institutional Review Board (IRB) UPMC Pittsburgh. PI: Ethan Rossi, PhD. Title: Laser Doppler Holography. Approval Date: 12/21/2022. Effective Date: 1/9/2023.

## Devices and software for clinical research

- Conception of reliable, robust, and low-maintenance clinical Doppler holography devices that ensure operator-independent measurements.
- Design, maintenance, and support of real-time (holovibes) and offline (holowaves) streamlined hologram image rendering software.
- Development of an analysis software (pulsewave) for quantitative biomarker assessment, targeting the evaluation of arterial pulse wave parameters, blood velocity, volume rate, arterial resistivity index, elastography, and blood rheology.
- All device and software initiatives are pursued with the intent to facilitate and bolster technology transfer to medical imaging companies and foster collaborations with academic institutions.

## Imaging patients

- 2018- : comprehensive study of ~600 patients with diverse eye pathologies and control subjects using Doppler holography. Clinical imaging with a prototype device at the Quinze-Vingts eye hospital.

- 2019- : acquisition and maintenance of a growing database of raw and rendered patient data at the Quinze-Vingts eye hospital (current size : ~300 TB).

# Summary note

## Overview of the development and milestones of Doppler holography for ophthalmology

This summary provides an overview of the progress and significant milestones achieved in the field of Doppler holography applied to ophthalmology. The technique has shown great potential in non-invasively visualizing blood flow patterns in ocular tissues, enhancing our understanding of various eye conditions and diseases. Starting with its initial conception to the latest breakthroughs, this overview highlights the key advancements that have shaped the use of Doppler holography as a valuable tool in ophthalmic research and clinical practice.

# Laser Doppler imaging, revisited

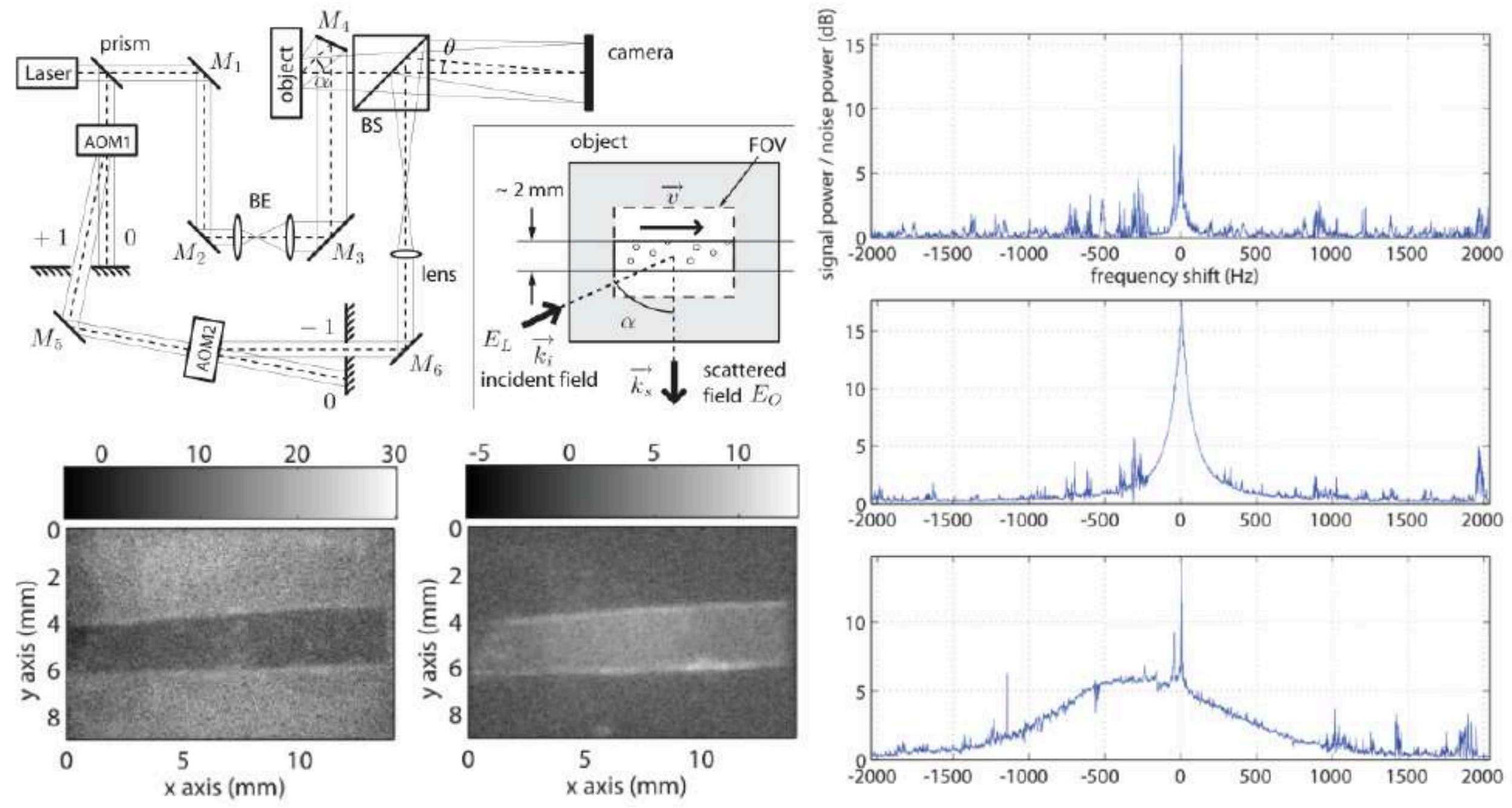

*Fig. 1 : **Top left** : Off-axis, frequency-shifting Doppler holography Setup. M1 to M6: mirrors; BS: beam splitter; BE: beam expander; AOM1 , AOM2: acousto-optic modulators (Bragg cells); -1, 0, +1 : diffraction orders. Inset : object and optical configuration. FOV: field of view. α: scattering angle. v : flow direction. ki: incident wave vector. ks: scattered wave vector, in the direction of the receiver. $E_L$: laser (incident) field. $E_O$: field scattered by the object. $E_{LO}$: reference (local oscillator) field. **Bottom left** : sample perfused with an estimated average velocity of 870 µm/s. Doppler maps at zero detuning frequencies of 0 and -8 Hz, displaying a contrast reversal between the signal in the channel and the static background. **Right** : spectrum in dB of the static background region (top), and in the channel region for null average velocity (center), and v = 870 µm/s (bottom).*

My research activity in Doppler imaging by digital holography began in 2005 within Michel Gross's group at the Kastler Brossel laboratory of the Ecole normale supérieure. The inaugural study introduced a digital holography detection scheme tailored for wide-field Doppler imaging, with a primary emphasis on characterizing flows having velocities around 1 mm/s in an expansive microfluidic channel [45]. In this exploration, we employed tunable, narrowband spectral measurements using time-averaged digital holography combined with a low frame rate camera. The temporal frequency analysis of optical field fluctuations was achieved by detuning the optical frequency of the reference wave. This was accomplished by sampling the beating of the reflected radiation against a frequency-shifted optical reference beam with the array detector of a conventional video-rate camera. Sequential changes in the detuning frequency allowed us to capture images of the local fluctuation spectrum of laser light backscattered by the fluidic channel. By leveraging spatiotemporal heterodyning (a fusion of off-axis and frequency-shifting detection), we achieved high detection sensitivity, particularly in the high heterodyne gain regime, where the optical power of the reference beam greatly surpasses that of the signal beam. This approach effectively minimizes parasitic interferometric contributions and enables the required sensitivity for imaging local optical fluctuation spectra beyond the camera bandwidth. This innovative method facilitates the acquisition of spectral data related to the dynamic state of local scatterers within the moving fluid, encapsulating the momentum transfer from dynamic light scattering events (Fig. 1). Importantly, this seminal research represents the first successful deployment of Doppler imaging by heterodyne off-axis digital holography.

## Laser Doppler imaging of microflow

A later study showcased the imaging of particle-seeded microflows, with velocities spanning from 0 to 50 µm/s, using time-averaged heterodyne, off-axis Doppler holography paired with a video-rate camera [46]. The resulting Doppler frequency distribution displayed asymmetry. Its first-order moment, expressed in Hz, provides a quantitative and directional measure of local particle velocity (Fig. 2). While direct image analysis techniques in microfluidic systems -such as fluid seeding and particle image velocimetry- yield more precise flow estimates for these low flow rates, this research confirms the potential application of Doppler holography in the quantitative characterization of microflows.

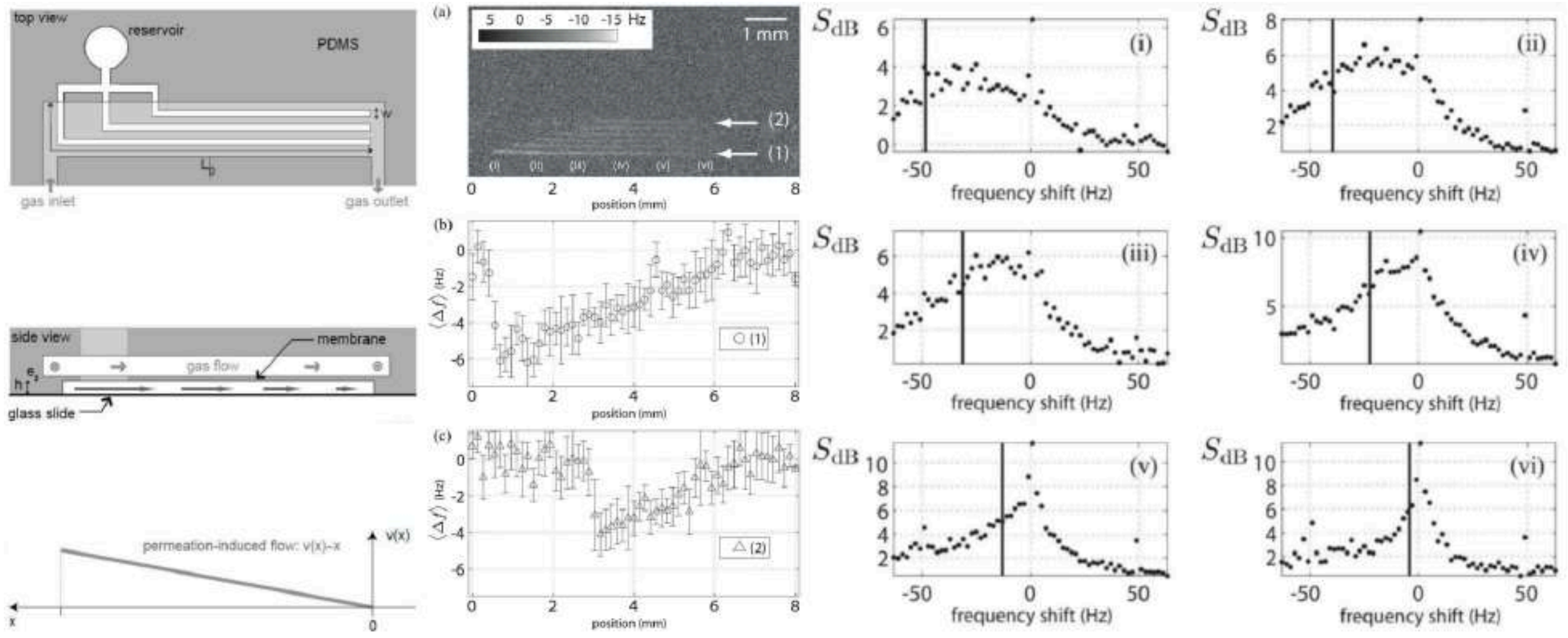

*Fig. 2 : **Left** : Sketch of the microfluidic device generating linear velocity fields via evaporation. This device is made of a thin PDMS membrane (thickness e ~ 10 µm) on glass. Typical dimensions are Lo ~ 1 cm, h ~ 20 µm. The permeation-induced flow is stationary, and varies linearly in space between 0 and a typical upper bound velocity of 50 µm/s (longest channel). **Center** : First moment of the Doppler spectrum, in Hertz. Map (a) and average signal plotted versus longitudinal position in channels 1 (b) and 2 (c), outlined in (a) . **Right** : Spectra (in dB) averaged six regions of interests outlined in (a); Vertical markers represent the expected average flow-induced frequency shift in each region of interest.*

## Frequency-domain wide-field laser Doppler in vivo imaging

After our successful demonstrations of Doppler imaging in microfluidic channels, a significant milestone was reached with the first-ever measurement of blood flow in the mouse brain using heterodyne (both off-axis and frequency-shifting) Doppler holography [47]. Fig. 3 presents a dorsal view of a living mouse's cranium, capturing the distinct Doppler signature of blood flow in the superficial layers of the brain. In the vessels highlighted by Doppler contrast, flow direction becomes indistinguishable due to the random scattering of light in the tissue. The local Doppler lineshape width, denoted in Hz and derived from the second-order moment of the observed Doppler frequency distribution, exhibits an increase in the superficial cerebral dorsal veins and arteries compared to the parenchymal regions. These findings underscore the potential of Doppler holography for non-invasive monitoring of blood flow in living beings, especially within superficial vascular networks. A primary limitation of this method is the requisite to sweep the detuning frequency, resulting in a total signal acquisition time spanning several minutes. This extended duration complicates the tracking of hemodynamics throughout a cardiac cycle.

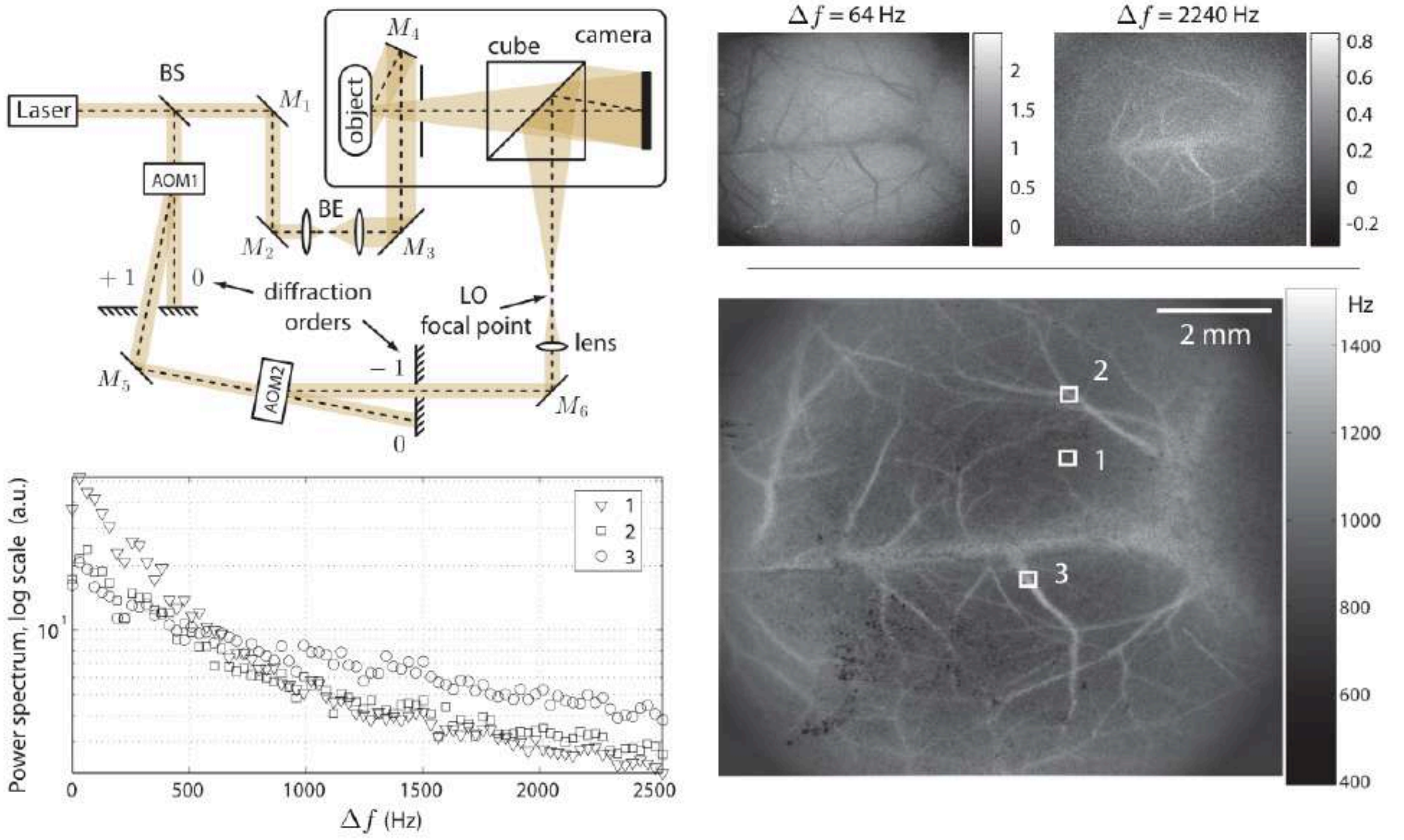

*Fig. 3 : **Top left** : Optical configuration : Off-axis, frequency-shifting Doppler holography. AOMs : Bragg cells. BS : beam splitter. BE : beam expander. M1 to M6 : mirrors. **Right** : Power maps of the Δf = 64 Hz and Δf = 2240 Hz frequency components of the scattered field (top, log. scale) and local Doppler frequency broadening, in Hertz (bottom). The image shows a dorsal view of the mouse cranium (anterior on the left, posterior on the right). The superficial dorsal venous system and some of the superficial cerebral arteries are visible. **Bottom left** : Spectra (log. scale) in the three regions of interest outlined in the image.*

## Cortical blood flow assessment with frequency-domain laser Doppler microscopy

Doppler holography (combining frequency-shifting and off-axis interferometry to enable tunable narrowband detection) can be employed to monitor cerebral blood flow alterations in mice, with the second-order moment of the measured frequency spectrum acting as an indicator of local cerebral blood flow velocity. A study [44] observed that this value declined notably, by 10% to 20%, between 10 and 20 minutes post catecholamine injection in specific superficial brain regions. Cerebral blood flow reverted to its pre-injection levels approximately 1 hour and 20 minutes after the injection. These observations align with prior in vivo and in vitro studies, which have indicated that catecholamines trigger vasoconstriction, contraction, a surge in perfusion pressure, and a subsequent reduction in blood flow. However, the study doesn't provide compelling evidence for the routine application of Doppler holography in its presented format for blood flow tracking because the rodent had to be held in a stereotactic frame, and measuring the Doppler spectra necessitated several minutes of data collection.

## Holographic laser Doppler ophthalmoscopy

Using near-infrared, tunable heterodyne Doppler holography, imaging of the eye of a living rat was demonstrated with a low frame rate camera in a pilot study [28]. To ensure minimal interference, the power of the near-infrared (785 nm) illumination beam was maintained at a low level, approximately 1 mW, across the entire eye fundus. A coverslip was affixed to a ring encircling the eye globe, supplemented by a transparent contact medium to filter-off unwanted light reflections. By adjusting the frequency of the reference wave, local optical field fluctuation spectra were recorded. The resulting wide-field maps of these optical fluctuation spectra, covering a frequency range from 10 Hz to 25 kHz, unveil angiographic contrasts within the retinal vascular tree (Fig. 4). The obtained signals display high consistency over prolonged periods, across repeated experiments, and between different animals. Such reliability hints at the method's potential in studying retinal and possibly choroidal vascular diseases. It's worth highlighting that signals near retinal vessels are believed to be significantly influenced by choroidal flow, a notion further corroborated by subsequent research. Within the retinal vessels, a broader Doppler spectrum is discernible compared to adjacent areas, implying that local retinal blood flow contributes to additional Doppler broadening. Nonetheless, time-averaged heterodyne Doppler holography, with a slow camera under off-axis interferometric conditions, exhibits less-than-ideal lateral and temporal resolution, hindering the detection of flow variations associated with the heartbeat.

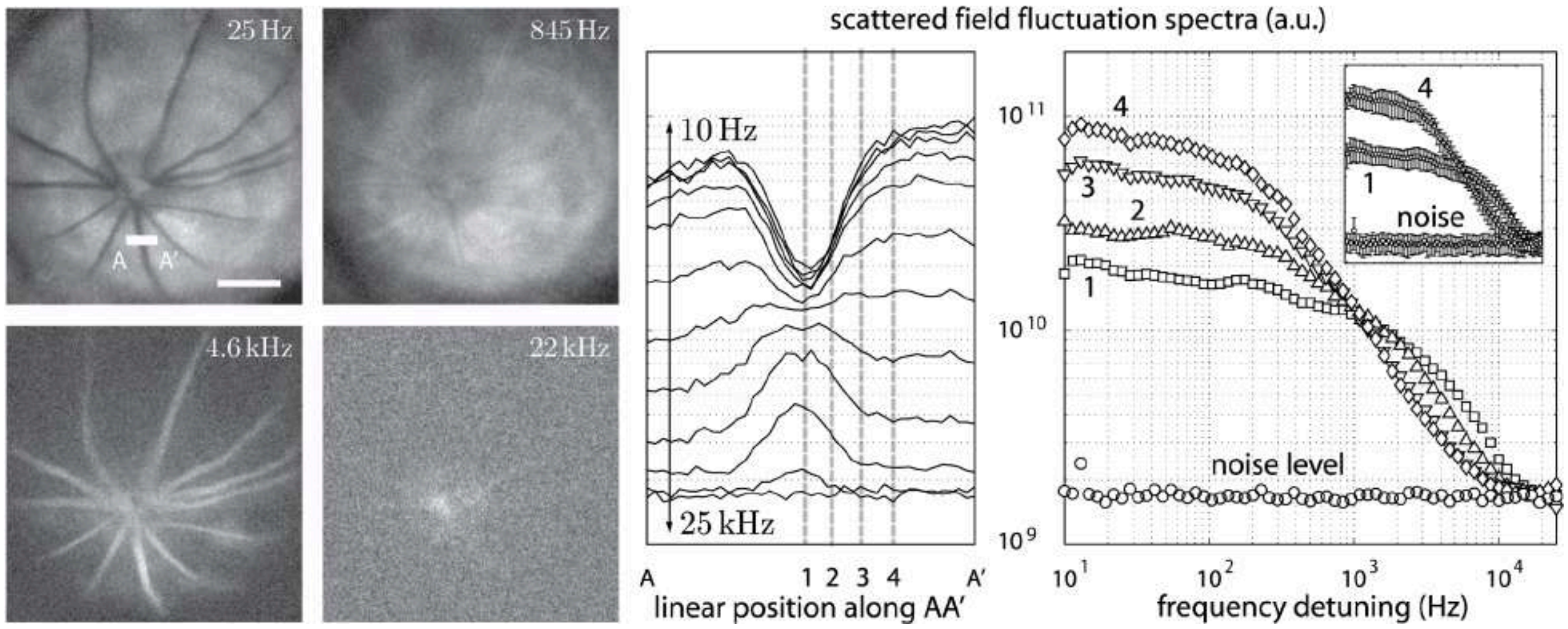

*Fig. 4 : **Left** : Doppler images of the eye fundus of a rat in logarithmic scale (white is for high signal). Frequency detuning : $\Delta\omega/(2\pi)$ = 25 Hz, 845 Hz, 4.6 kHz, 22 kHz. [AA'] indicates the measurement region of the lines reported in the central plot. Scale bar, 1 mm. **Center** : Spectral components against position along [AA']; These traces correspond to logarithmically-spaced detuning frequencies from 10 Hz to 25 kHz. **Right** : Spectral components against frequency detuning at positions 1, 2, 3, 4 from the center of the vessel to its periphery, suggesting that the Doppler broadening increases with flow in the retinal vessels.*

## Holographic laser Doppler imaging of microvascular blood flow

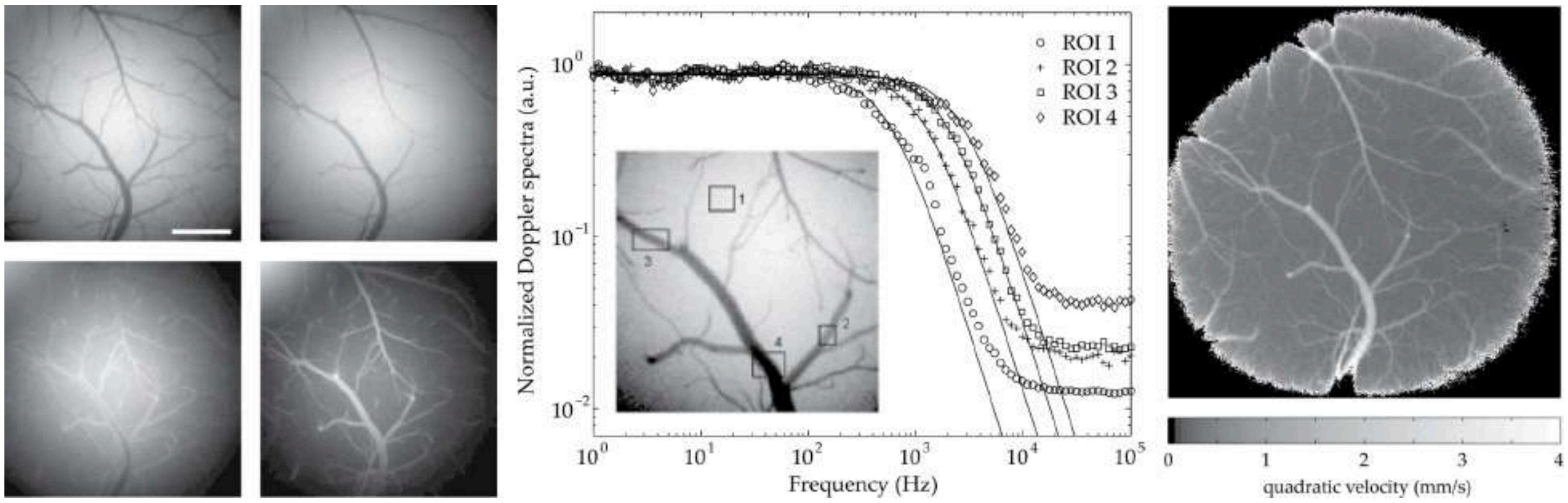

*Fig. 5 : **Left** : Doppler images of the mouse cerebral cortex at different frequency shifts : 9 Hz (top left), 167 Hz (top right), 1.7 kHz (bottom left), 5.5 kHz (bottom right). Arbitrary logarithmic gray scale. Scale bar : 1 mm..*

***Center** : Normalized first-order power spectra averaged in the regions of interest (ROI) 1 to 4 (dots correspond to measurements, continuous lines are the result of fitting measured values with a Lorentzian lineshape). Insert: Doppler map displaying the 4 regions of interest. **Right** : Map of the local quadratic mean blood flow velocity derived from the Doppler images.*

Local superficial blood flow in biological tissue can be assessed using a basic diffusing wave spectroscopy model in a backscattering configuration, from Doppler holography [15]. By analyzing the optical fluctuation spectra gathered by narrowband heterodyne Doppler holography within a frequency range of 0 Hz to 100 kHz, one can evaluate the quantitative quadratic mean blood flow velocity within the superficial vascular network of both the rodent retina and exposed brain (Fig. 5). The lateral spatial resolution of approximately 10 microns allows visualization of small superficial vessels and arteries. Observations suggest that :

- Near-infrared laser light fluctuation spectra from the cerebral cortex of a mouse, the eye fundus of a rat, and a sizable microfluidic channel display a Lorentzian lineshape. This lineshape's width augments in line with local flow velocity.
- Under controlled flow rates within a fluidic channel, the width of the optical fluctuation spectrum lineshape scales linearly with fluid flow velocity, ranging from 100 µm/s to 10 mm/s (Fig. 6).
- To ensure model accuracy, a global background frequency broadening was incorporated to accommodate the physiological movements of the retina. Assessing blood flow in the superficial retinal vasculature operates under the presumption that discernible vessels induce added Doppler broadening relative to the surrounding tissue. This disparity is ascertained by contrasting the width of the local Doppler lineshape with the global Doppler lineshape sourced from background tissue.

It can be highlighted that: (i) In retinal vessels, the Doppler lineshape broadening may more so reflect the influence of choroidal blood flow rather than eye motion in velocity estimations. (ii) The velocity estimation for blood flow within the superficial vessels of the cerebral cortex [15] (Fig. 5) could have been enhanced by adopting a similar model. This model would take into account a global background Doppler broadening stemming from the densely vascularized brain parenchyma, as it is the case for the choroid.

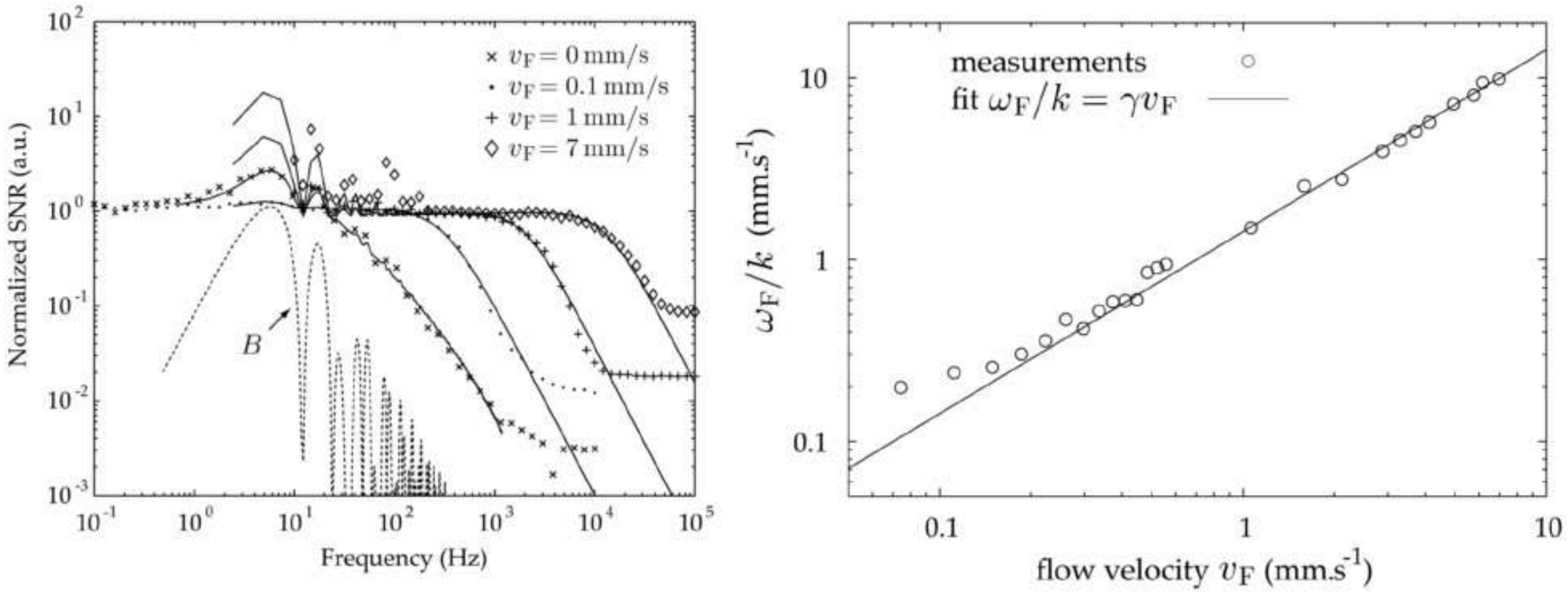

*Fig. 6 : **Left** : Normalized spectral lines (a.u., symbols) versus detuning frequency, in Hz. The average perfusion velocity in the tube ranges from 0 mm/s to 7 mm/s. The measured lines are fitted against a Lorentzian lineshape for non-null convective motion of the scatterers, plotted as continuous black lines. The apparatus lineshape is plotted as a dotted line. **Right** : Doppler broadening estimated from the width of the Lorentzian versus average perfusion quadratic velocity, in mm/s. The Doppler linewidth broadening scales up linearly with the velocity for convective motion within a range of 0.1 mm/s to 10 mm/s.*

## High-speed wave-mixing laser Doppler imaging in vivo

Broadband detection of the optical fluctuation spectrum using a high-speed camera and Fourier-transform analysis greatly improves Doppler holography [37]. Time-averaging recording conditions achieved with a slow camera create a narrowband bandpass filter that requires sequential adjustments through detuning of the optical frequency of the reference beam to sample the optical fluctuation spectrum. In contrast, high-speed broadband Doppler holography employs a 2 kHz frame rate camera, enables the visualization of partial Doppler contrast variations between the superficial vessels of the cortex and the parenchyma in a rodent, through the skull, under near-infrared laser light illumination (Fig. 7). The introduction of broadband detection and Fourier-transform analysis of the optical fluctuation spectrum using a high-speed camera were crucial for subsequent assessments of blood flow variations during cardiac cycles.

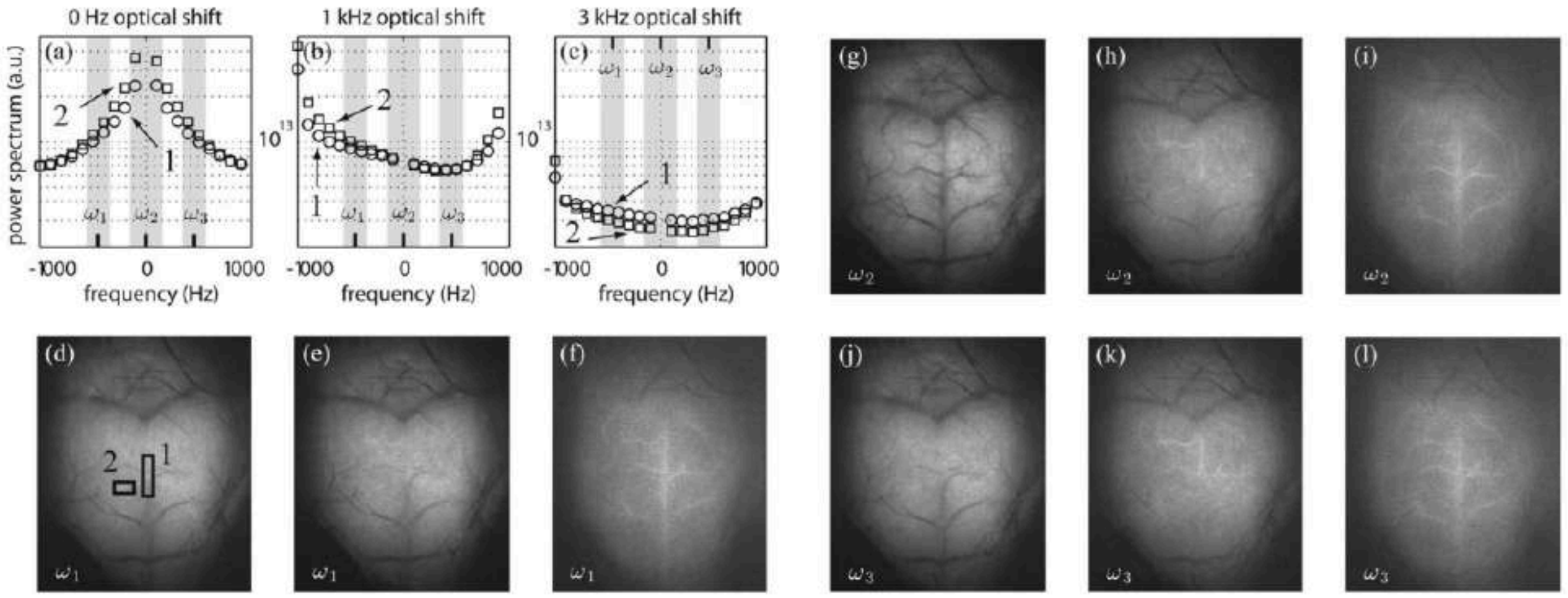

*Fig. 7 : Doppler holography of the cranium of a mouse in off-axis, frequency shifting configuration with a 2 kHz frame rate camera. (a)−(c) Spectra and (d)−(l) image components, in logarithmic scale. Fluctuation Spectra are measured in the regions of interest 1 and 2, sketched in (d). Results obtained at three optical frequency shifts: (a), (d), (g), (j) detuning frequency : 0 Hz; (b), (e), (h), (k) detuning frequency : 1000 Hz; (c), (f), (i), (l) detuning frequency : 3000 Hz. Rows : images of three frequency bands fromFourier analysis [displayed in gray in (a)−(c)]: (d)−(f) $\omega_1$=-450 Hz, (g)−(i) $\omega_2$=0 Hz, (j)−(l) $\omega_3$= +450 Hz. Contrast variations between the vessels and background for both optical frequency detuning and exploration of the computed Doppler spectrum suggest the suitability of  broadband Doppler holography with ultra-fast cameras for blood flow imaging, in-vivo.*

## Pulsatile microvascular blood flow imaging by short-time Fourier transform analysis of ultrafast laser holographic interferometry

Pulsatile microvascular blood flow can be revealed in the exposed mouse cortex using high-speed off-axis digital holography with a 50,000 Hz frame rate camera [c1]. The technique involves Fresnel transformation for hologram rendering and short-time Fourier analysis to extract localized Doppler contrasts from blood flow. A laser of 532 nm wavelength illuminates the exposed cortex via a cranial window, while cross-polarized laser detection filters out specular reflections. These reconstructed off-axis holograms, spanning a 4x4 mm cortical area, achieve a resolution of 10 µm.

Through short-time Fourier transforms, time-frequency maps are produced. Here, the width of the frequency lineshape serves as an indicator of average perfusion, revealing pulsatile dynamics in arteries across cardiac cycles. Lower Doppler frequencies unmask finer vessels, whereas higher frequencies underscore more prominent vascular structures with larger expected flow velocity. Composite color images seamlessly integrate these varying scales, as shown in Fig. 8.

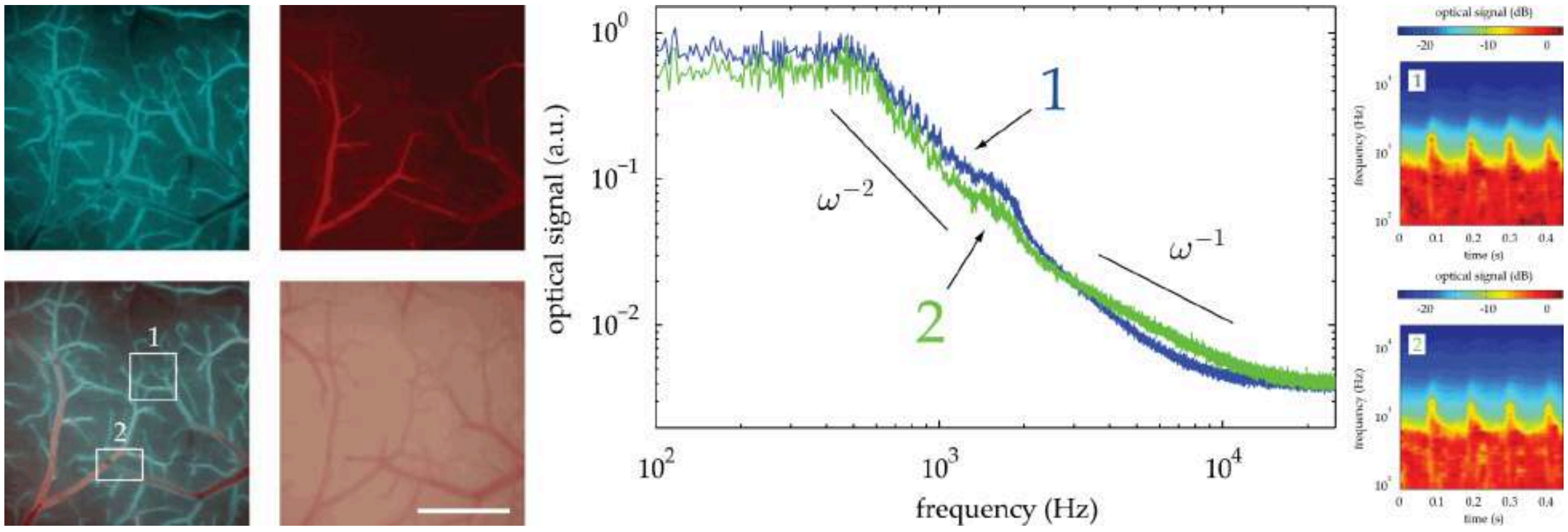

*Fig. 8 : **Left** : Images of the exposed cortex of a mouse. root-mean-square frequency calculated in the bands [2.3 kHz, 5 kHz] (cyan) and [5 kHz, 25 kHz] (red). Composite color Doppler image (movie), and white-light microscope image of the same area. Scale bar : 1 mm. **Center** : Time-averaged optical signal S (top) versus frequency (movie). Both quantities are averaged spatially in the region 1 (blue) and region 2 (green) depicted in the composite image. **Right** : spectrograms in region 1 and region 2, displaying arterial pulse waves.*

Doppler holography with a high-speed camera allows for the visualization of localized cortical blood flow contrasts at spatial and temporal scales pertinent to microvascular perfusion. This method captures cardiac pulsations in small superficial vessels, delivering discernible hemodynamic contrasts via spectrograms calculated by time-frequency Fourier analysis. Significantly, the spectrogram within the densely vascularized brain parenchyma also displays a pulsatile variation synchronized with each heartbeat, which suggests an important contribution of the background signal to the Doppler lineshape measured in superficial vessels. While this demonstration is focused on the exposed cortex, the pilot study lays the groundwork for exploring retinal microcirculation. This is especially promising given the success of high-speed blood flow contrast measurements using short-time Fourier analysis, boasting a temporal resolution of approximately 20 ms.

## Holographic laser Doppler imaging of pulsatile blood flow

Heterodyne holographic interferometry offers the capability for wide-field imaging of pulsatile motion induced by blood flow, as evidenced by real-time observations on a healthy volunteer's thumb [14]. We showcased optical Doppler image rendering using green laser light through a frequency-shifted Mach-Zehnder interferometer set in an off-axis configuration. The captured optical signal can be correlated to local, instantaneous, out-of-plane skin motion, with velocities reaching several hundred micrometers per second. This data aligns with blood pulse patterns observed via plethysmography during an occlusion-reperfusion experiment (Fig. 9).

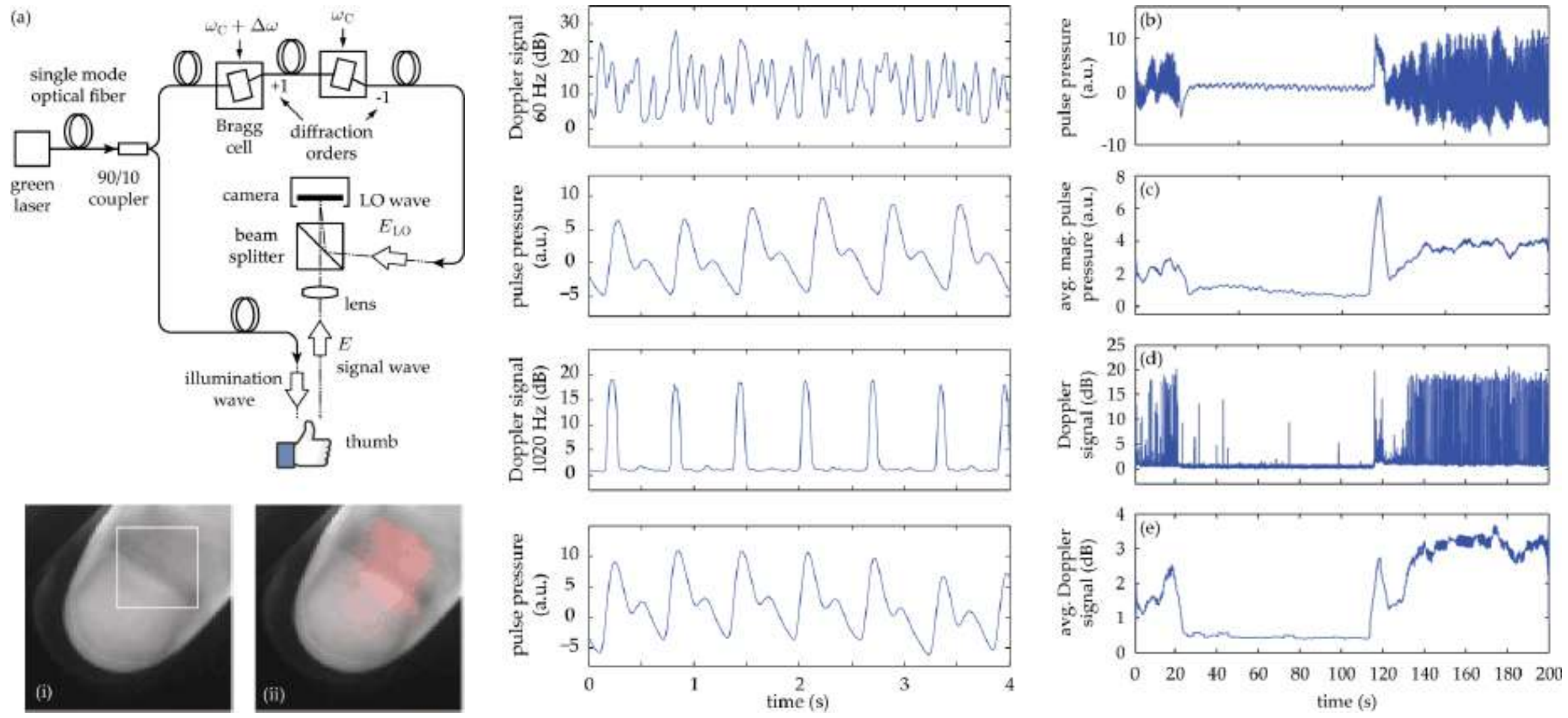

*Fig. 9 : **Left** : Sketch of the fibered Mach-Zehnder heterodyne holographic interferometer used for Doppler holography of the thumb (a). The main laser beam is split into two channels. In the object channel, the optical field is backscattered by the skin. The reference field is frequency-shifted by two Bragg cells by the quantity $\Delta\omega$. A standard camera records interferograms of the diffracted fields beating against each other. Reconstructed images of the measured thumb (movie). **Center** : Doppler signal, averaged in the central part of the image of the thumb (i), and concurrent pulse pressure, versus time. Detuning frequencies: $\Delta\omega/(2\pi)$ = 60 Hz, and $\Delta\omega/(2\pi)$ = 1020 Hz. The synchronization accuracy of holographic and standard blood volume measurements is on the order of ±0.5 s. **Right** : Time traces of indicators monitored during a hypoperfusion experiment. plethysmogram [(b) raw data, (c) 5 s time-average of the magnitude], holographic laser Doppler signal recorded at 1020 Hz and averaged spatially in the nail area [(d) raw data, (e) 5 s time-average]. Linked media shows a time-lapse sequence of the composite Doppler image of the thumb (movie).*

In that experimental demonstration, off-axis interferograms were recorded at a frame rate of 60 Hz. Short-time Fourier analysis extracts localized motion contrasts linked to the cardiac pulse wave. A 532 nm laser illuminates the thumb area of 40x40 mm, and backscattered light interferes with a reference beam frequency-shifted by acousto-optic modulators, allowing for heterodyne downconversion of Doppler shifts into the camera bandwidth. This real-time demonstration of imaging cardiac-induced pulsatile signals in the thumb, using standard 60 Hz camera with hologram rendering and Fourier analysis, lays the groundwork for future developments in high bitrate hologram rendering software.

## High speed optical holography of retinal blood flow

Off-axis Doppler holography with near-infrared laser light permits the visualization of pulsatile retinal blood flow in pigmented rats. The approach uses interferogram recording at a frame rate of 39,000 Hz with a high-throughput camera [11], allowing imaging of large retinal vessels with exceptional temporal resolution (~6.5 ms). Digital holograms are reconstructed by Fresnel transformation, and Doppler flow contrasts are extracted by short-time Fourier analysis. By utilizing cross-polarized illumination and collection, the moving red blood cells are isolated from static tissue, enhancing the clarity of the blood flow signals. The reconstructed holograms cover a 3.3 mm x 3.3 mm field of view with an 8µm resolution. Importantly, the second moment of the Doppler spectrum reveals endoluminal pulsatile flow in arteries over cardiac cycles, exhibiting distinct waveforms in retinal arteries and veins (Fig. 10). Composite color images combining different moments effectively highlight vascular anatomy and flow patterns. High-speed Doppler holography successfully provides detailed and substantial retinal pulsatile blood flow contrasts in rodents.

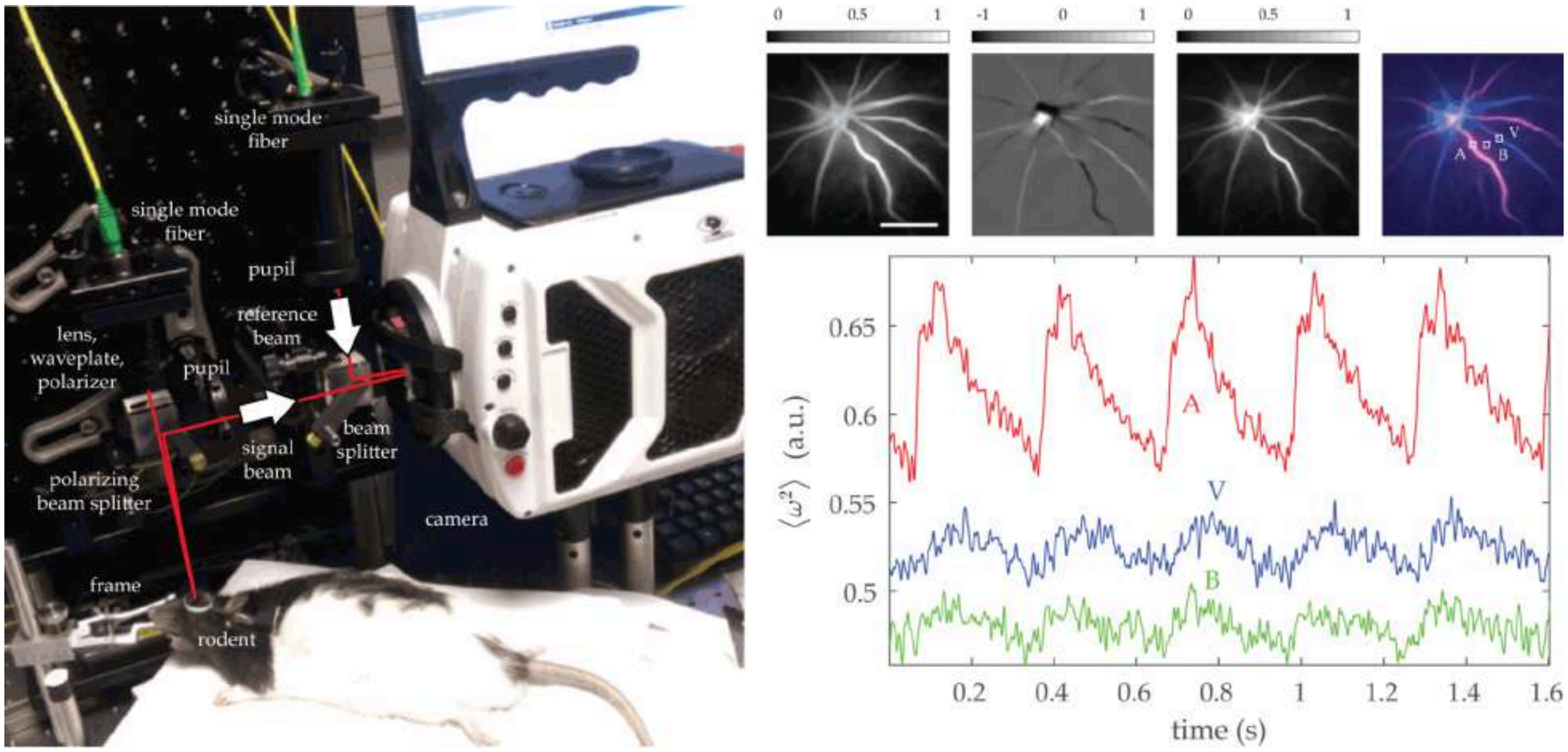

*Fig. 10 : **Left** : Picture of the fibered Mach-Zehnder holographic interferometer used for retinal imaging of the eye fundus of a rat. Off-axis interferograms of near-infrared radiation are recorded by the sensor array of a high speed camera. **Top right** : Zeroth, first, and second moments of the envelope of the local fluctuation spectrum. Composite color image of the zeroth and second moment ([movie](movie)). Scale bar ~ 1 mm. Color scales : arbitrary units (a.u.). **Bottom Right** : Second moment moment of the optical fluctuation spectrum versus time, averaged in the regions of interest A (retinal artery), V (retinal vein), B (background), depicted in the color image.*

## In vivo laser Doppler holography of the human retina

The first blood flow imaging in the human retina using Doppler holography [9] is based on on-axis digital holography with a ultra-high speed camera. Dual imaging channels—comprising both slow and fast cameras—are employed for real-time monitoring and offline analyses, respectively. A 785 nm laser diode illuminates the retina, and cross-polarization filters out parasitic light components. Inline digital interferograms are recorded simultaneously on the slow camera for real-time monitoring and on the fast camera up to 75 kHz for offline processing. By analyzing the beat frequency spectrum through short-time Fourier transform, power Doppler images are generated, revealing blood flow contrast in retinal vessels. This technique enables observation of cardiac-cycle pulsatility in arteries with high temporal resolution (Fig. 11). Doppler holography offers unique advantages, including retinal blood flow contrasts, access to lateral motion, high temporal resolution, and full-field imaging. It effectively detects Doppler shifts at low light levels, providing new insights into retinal blood flow physiology and dynamics. The successful demonstration of perfusion imaging in a living human retina heralds a new era for non-invasive evaluations of ocular blood circulation.

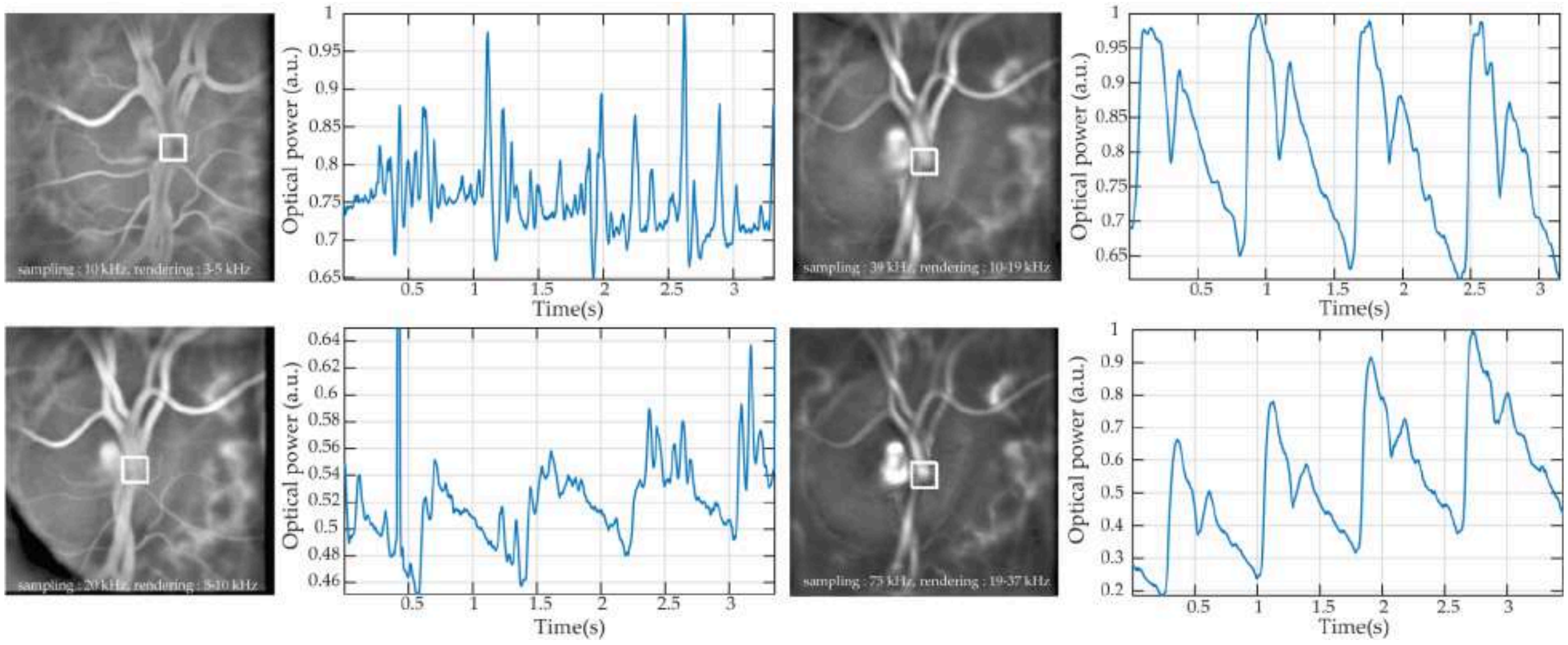

*Fig. 11 : Doppler holography in the central retinal artery. For each sampling frequency (10, 20, 39, and 75 kHz), power Doppler images are spatially averaged over the depicted regions of interest on the left hand side and the result is displayed in the associated plot. The Fourier transform parameters used for each acquisition have been chosen to have a short-time window of 13 ms for all acquisitions. Camera sampling frequency : 10 kHz (top left), 20 kHz (bottom left), 39 kHz (top right), 75 kHz (bottom right). Linked media shows power Dopler images as a function of the frequency, for a camera sampling frequency of 75 kHz (movie).*

## Choroidal vasculature imaging with laser Doppler holography

The choroid, a highly vascularized tissue supplying the retinal pigment epithelium and photoreceptors, is gaining increasing interest in retinal disease research. However, investigating its anatomy and flow characteristics remains challenging. Doppler holography can furnish high-contrast imaging of the choroidal vessels in humans, exhibiting a prowess on par with state-of-the-art methods such as indocyanine green angiography (ICG-A) and optical coherence tomography [8]. Doppler holography effectively distinguishes choroidal arteries and veins through power Doppler spectral analysis (Fig. 12). It reveals previously unseen choroidal arteries around the optic nerve and para-optic short posterior ciliary arteries branching into the choroid. In the posterior pole, more submacular choroidal arteries are visualized compared to ICG-A, and Doppler holography images match the early arterial phase of ICG-A. The technique's wide-field imaging capability offers new insights into choroidal anatomy and physiology, including vortex veins outside the posterior pole. Overall, Doppler holography noninvasively provides valuable information about the choroidal vascular network, enhancing our understanding of this complex tissue.

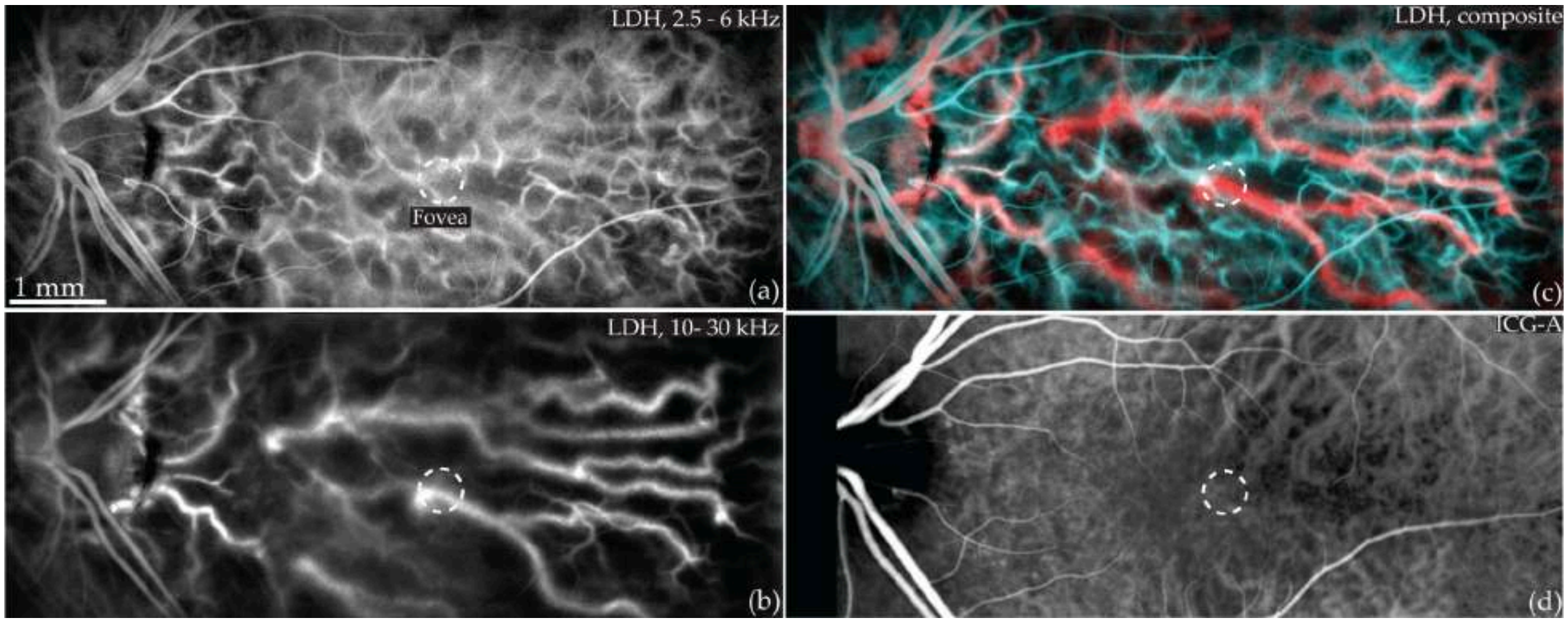

*Fig. 12 : Power Doppler images and indocyanine green angiography (ICG-A). (a) Low frequency (2.5-6 kHz) power Doppler reveals small flows. (b) High frequency (10-30 kHz) power Doppler reveals vessels with larger flows. Linked media displays a frequency sweep of the Doppler image (movie). (c) Composite color image of (a) and (b), encoded in the cyan and red channels, respectively. (d) ICG angiography of the same region.*

## Waveform analysis of human retinal and choroidal blood flow with laser Doppler holography

The capabilities of Doppler holography in imaging the choroidal vasculature in the human eye with high contrast and resolution are comparable and complementary to indocyanine green angiography (ICG-A), because Doppler holography provides sensitivity to blood flow and allows differentiation between choroidal arteries and veins [7]. It reveals choroidal arteries around the optic nerve and more submacular choroidal arteries than ICG-A in the posterior pole. The technique can discriminate vessels based on flow Doppler frequencies, revealing arterioles and venules at low frequencies and large choroidal arteries at high frequencies, offering arteriovenous contrast. Doppler holography also captures vortex veins outside the posterior pole with similar resolution to ICG-A. Overall, Doppler holography offers new insights into the anatomy and physiology of the choroidal vascular network, enabling non-invasive wide-field imaging of the choroid, which is not easily observable with other modalities. The introduction of Doppler holography as a full-field imaging technique for measuring blood flow in the retina and choroid has provided unrivaled temporal resolution.

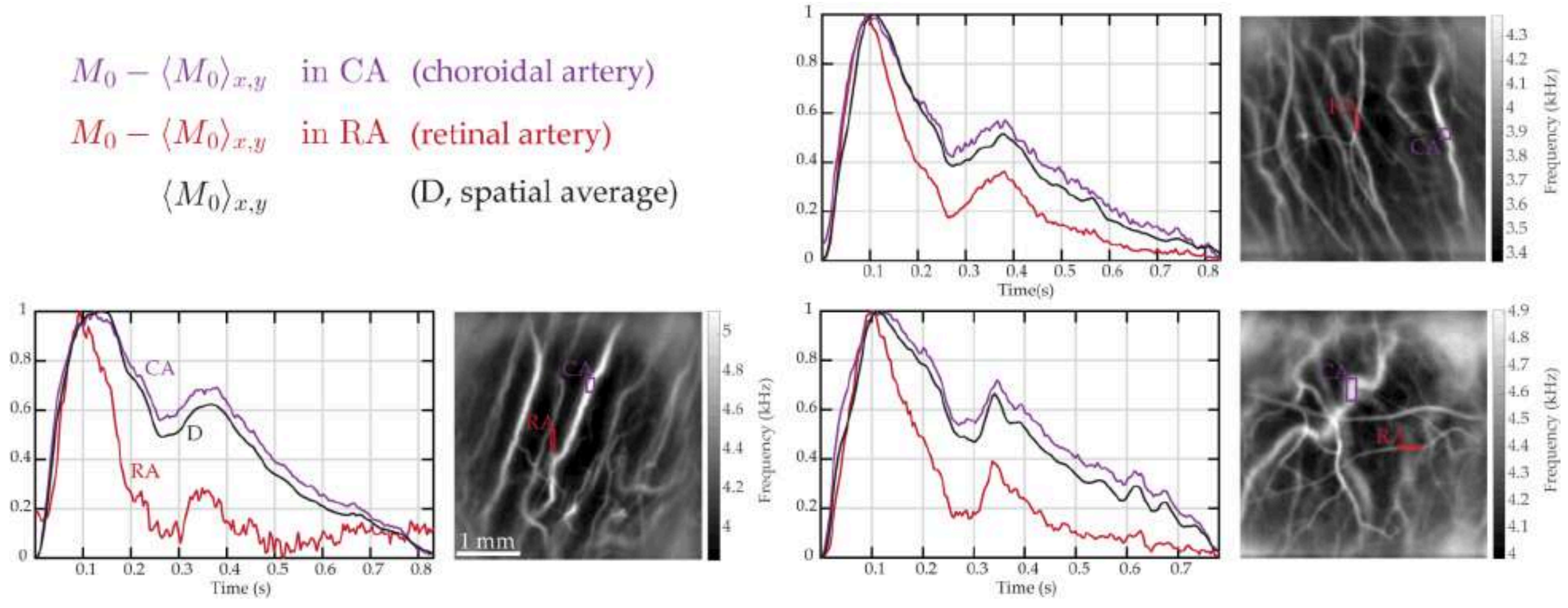

*Fig. 13 : Retinal and choroidal arterial waveforms compared to the spatial average in three examples in peripheral regions. Left column: normalized waveform of power Doppler corrected from the spatial average in a retinal and a choroidal artery ('RA' in red, and 'CA' in purple), and normalized waveform of the Doppler signal spatially averaged over the entire image ('D' in black). The dominant signal waveform has an arterial waveform, which strongly resembles that of choroidal arteries. These results suggest that subtracting the spatially averaged baseline signal to reveal local Doppler signals in visible arteries filters the strong spurious contribution of the Doppler signal from deep choroidal vessels.*

The power Doppler signal can reveal blood flow waveforms in vessels in the posterior pole of the human eye. Distinct flow behaviors are found in retinal arteries and veins with seemingly interrelated waveforms. Arterial flow waveforms in the retina and choroid are found to be synchronous and similar, albeit with lower pulsatility in choroidal arteries (Fig. 13). This suggests that the choroid is the primary contributor to the high-frequency laser Doppler signal in all reported fundus areas. However, its influence can be mitigated by subtracting the spatially averaged baseline signal. This overall signal is likely a result of light backscattering by the choroid through the intermediate layers, with a potential contribution to the Doppler broadening from the choroidal precapillary sphincter or choriocapillaris. This could explain why the spatially averaged signal reaches its systolic peak later than the retinal arteries.

## Spatio-temporal filtering in laser Doppler holography for retinal blood flow imaging

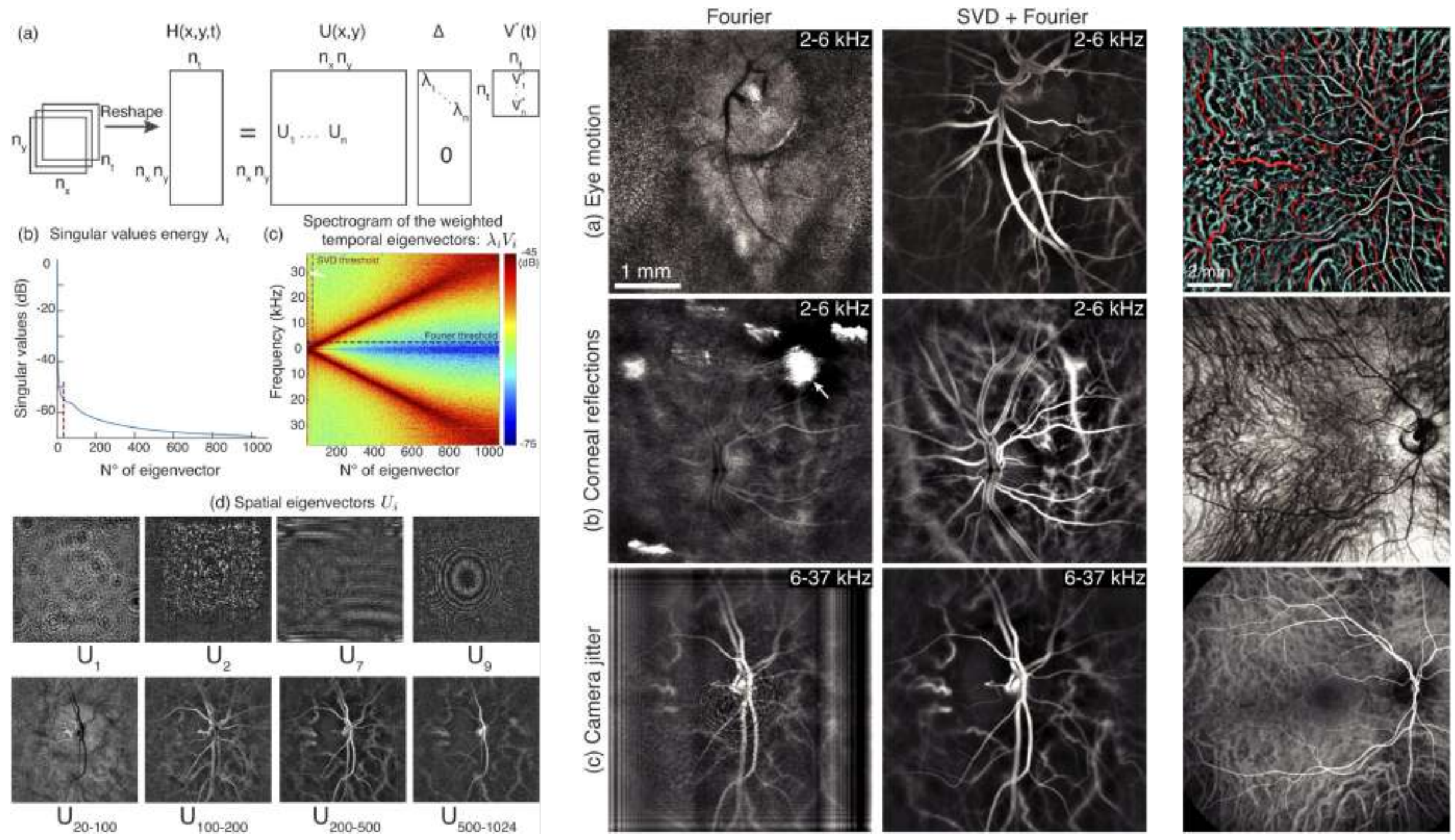

*Fig. 14 : **Left** : Singular value decomposition (SVD) for Doppler holography. (a) The 3D matrix of complex-valued holograms is reshaped into a 2D space-time matrix, and decomposed in a product of 3 matrices : the spatial and temporal eigenvectors, and the diagonal matrix of singular values. (b) Ordered singular values in dB. (c) Fourier transform magnitude of the temporal eigenvectors weighted by singular values. (d) spatial eigenvectors: the first vectors show spurious clutter whereas vectors associated with singular values of lower energy reveal blood flow and retinal features. **Center columns** : standard vs. SVD-filtered power Doppler images in the presence of: (a)*

*retinal motion (movie), (b) corneal reflections, and (c) camera jitter.* ***Right column*** *: Imaging the choroid with Doppler holography, OCT (Plex Elite 9000, Zeiss), and ICG-angiography (Spectralis, Heidelberg). The SVD-filtered low frequency component is critical to reveal choroidal veins.*

A statistical filtering approach based on singular value decomposition (SVD) effectively improves the capacity of Doppler holography to detect slower blood flow in the retina and choroid [6]. Doppler holography measures blood flow with high temporal resolution, but struggles with low Doppler frequency shifts corrupted by eye motion. Singular value decomposition filtering contributes to extracting a high-quality backscattered optical wave signal from the suboptimal on-axis interferometry configuration. It diminishes low-frequency spurious interferometric signal contributions and allows the measurement of high quality optical wavefields, facilitating precise and enhanced Doppler imaging of the retina. The proposed SVD filtering effectively separates blood flow signals from eye motion artifacts, enhancing the detection of smaller vessels and those with reduced flow (Fig. 14). The technique is demonstrated on retinal vein occlusion and choroidal vasculature imaging, providing superior visualization compared to conventional Fourier filtering, optical coherence tomography, and indocyanine green angiography. SVD filtering expands the utility of Doppler holography for studying various vascular diseases in the eye.

## Real-time digital holography of the retina by principal component analysis

Real-time digital holography can be done for high-quality imaging of the human retina using principal component analysis (PCA) for temporal demodulation. Inline digital holography for Doppler imaging is challenging due to low light conditions and interference from self-beating and twin images, resulting in subpar image quality with standard Fourier filtering on lower (less than ~5 kHz, typically) frame rate cameras. To address this, we implemented PCA on short temporal windows of consecutive digital holograms, identifying principal components to filter out spurious interferometric contributions [p1]. The setup uses a 785 nm laser to illuminate the retina, and cross-polarized backscattered light interferes with a reference beam. Real-time rendering of 1024x1024 pixel inline digital holograms is achieved at 500 Hz using Fresnel transformation on a GPU. The PCA approach proves superior to Fourier methods in filtering out interference, resulting in higher-quality retinal images (Fig.

15). Real-time PCA demodulation and visualization are achieved up to 20 Hz using commodity hardware. PCA effectively filters out interference adaptively by retaining lower-order eigenvectors carrying the blood flow signal of interest. This enables practical implementation of digital holography for retinal imaging and diagnostics in clinical settings, providing high-quality images with improved temporal demodulation.

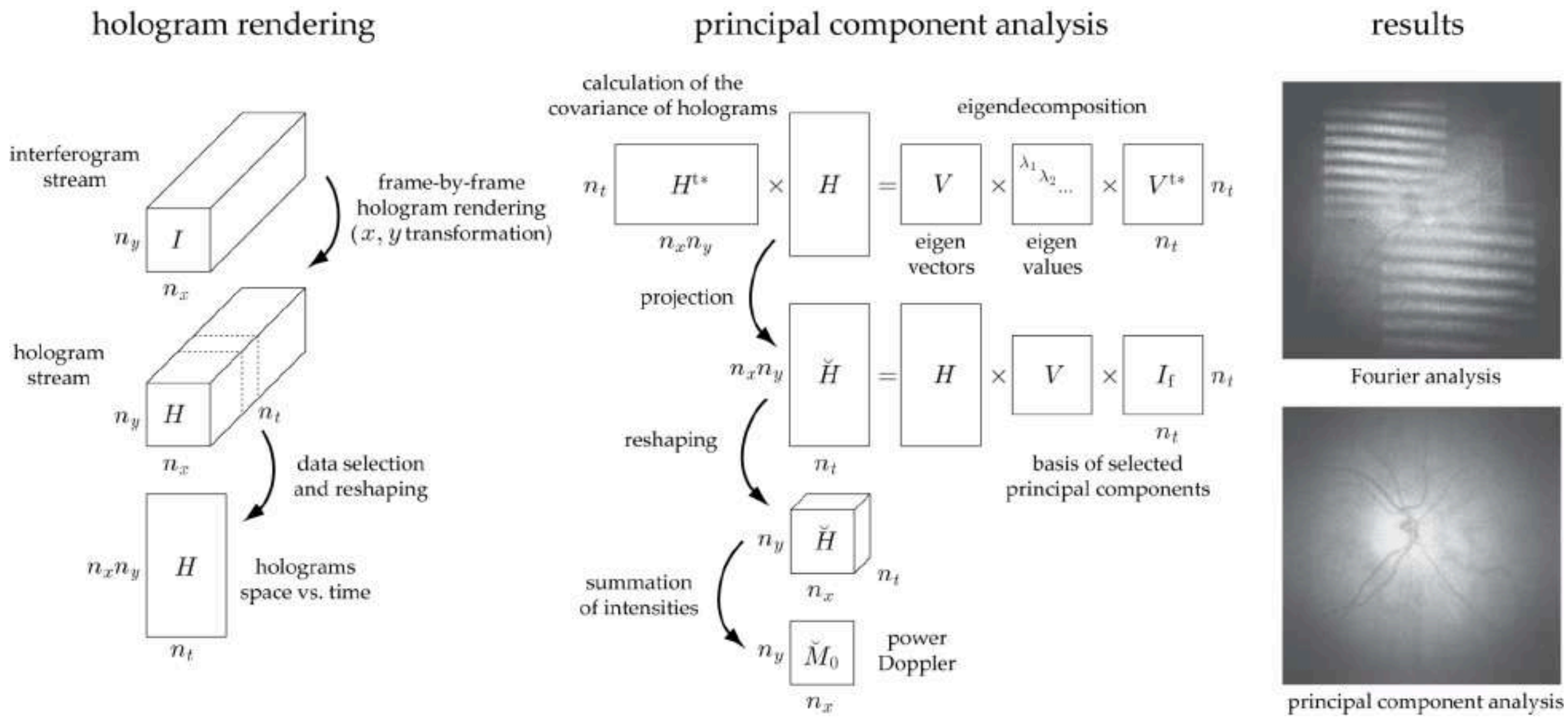

*Fig. 15 : **Left** : Flowchart of the data processing steps for temporal demodulation of retinal holograms by principal component analysis for real-time image rendering. **Right** : Real-time visualization of inline digital holograms of a human retina from an input stream of 16-bit, 1024-by-1024-pixel interferograms at a rate of 500 frames per second. The temporal signal demodulation of 32 consecutive holograms stacks is done by projection onto a Fourier basis (top) and onto a basis derived from eigendecomposition of the matrix of time-lagged covariance of consecutive holograms (bottom), at a rate of 20 Hz (movie). Principal component analysis effectively filters out spurious image features adaptively and reveals the blood flow signal of interest.*

# Reverse contrast laser Doppler holography for lower frame rate retinal and choroidal blood flow imaging

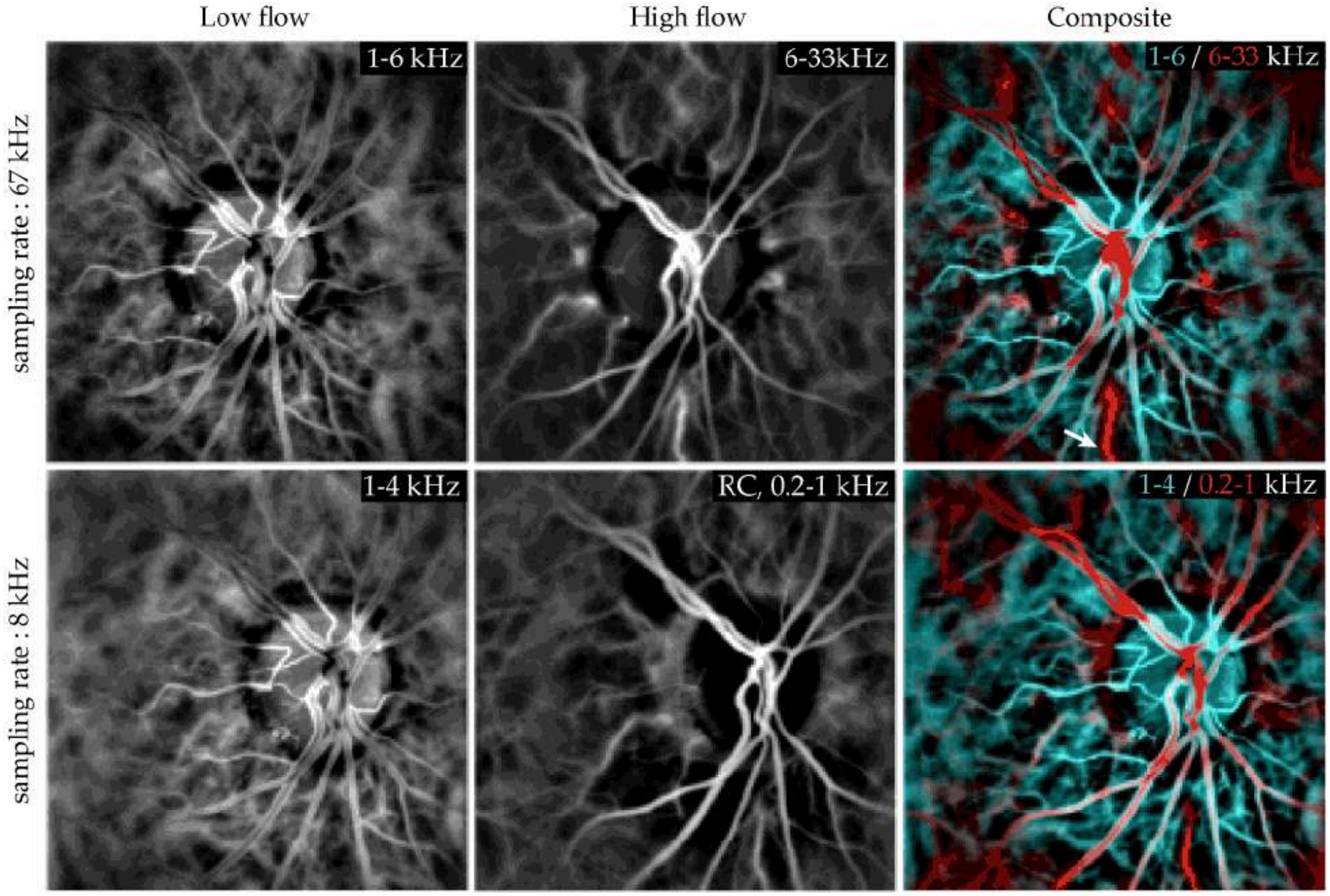

*Fig. 16 : Doppler images revealing low and high blood flow, obtained from 67 kHz (top row) and 8 kHz (bottom row) frame rate recordings.* ***Top row*** *: Power Doppler in the 1-6 kHz (left) and 6-33 kHz (center) frequency band; composite image (right).* ***bottom row*** *: Power Doppler in the 1-4 kHz (left) and 0.2-1 kHz (center) frequency band. The displayed contrast of the 0.2-1 kHz power Doppler is inverted (RC). composite image (right). Linked media shows composite color movies computed from sampled data at 67 kHz (movie) and 8 kHz (movie) in the same eye.*

Reverse-contrast Doppler holography achieves good visual results of pulsatile blood flow contrasts in retinal arteries at a sampling rate lower than 10 kHz [5]. The reverse-contrast Doppler holography approach was introduced as an attempt to enable blood flow imaging at much lower camera frame rates than standard high-speed Doppler holography which requires an ultrafast (about 75 kHz frame rate) camera. This method leverages the energy conservation of the Doppler spectrum under the hypothesis that the broadening of the spectral lineshape beyond the camera bandwidth provokes a loss of energy at low frequency. By inverting the contrast of low frequency power Doppler images, reverse-contrast Doppler holography effectively reveals fast blood flow components beyond the camera's conventional detection bandwidth. The reverse-contrast approach allows to obtain relevant blood flow variations and color composite power Doppler images at camera frame rates as low as a few kHz (Fig. 16). The technique is validated, demonstrating successful imaging of pathology induced flow changes and offering a practical clinical imaging method using slower cameras. It is important to note that this technique is made possible thanks to SVD filtering that enables the rendering of good quality Doppler images at low frequencies. Overall, reverse-contrast Doppler holography enables dynamic blood flow imaging at significantly reduced camera speeds compared to previous demonstrations. This is pivotal for the development of a first generation of clinical-grade Doppler holography instruments.

## Laser Doppler holography of the anterior segment for blood flow imaging, eye tracking, and transparency assessment

Without altering the Doppler holographic imaging scheme, digital holograms can be rendered for the anterior segment of the eye. Effective imaging is achieved for blood flow in the bulbar conjunctiva, episclera, and corneal neovascularization [4]. Simultaneous holographic imaging of the anterior and posterior segments can be achieved by adjusting the numerical propagation distance (Fig. 17). This feature allows tracking the movements of the retina and pupil with high temporal resolution. Additionally, light backscattered by the retina can be utilized for retro-illumination of the anterior segment, revealing opacities caused by absorption or diffusion in the cornea and eye lens. Digital holography offers multifaceted anterior segment imaging possibilities, enabling dynamic measurements and detailed visualization of various eye structures. The technique demonstrated higher contrast and a wider field of view compared to slit lamp imaging for assessing lens and corneal transparency, making it a valuable approach in ophthalmic research.

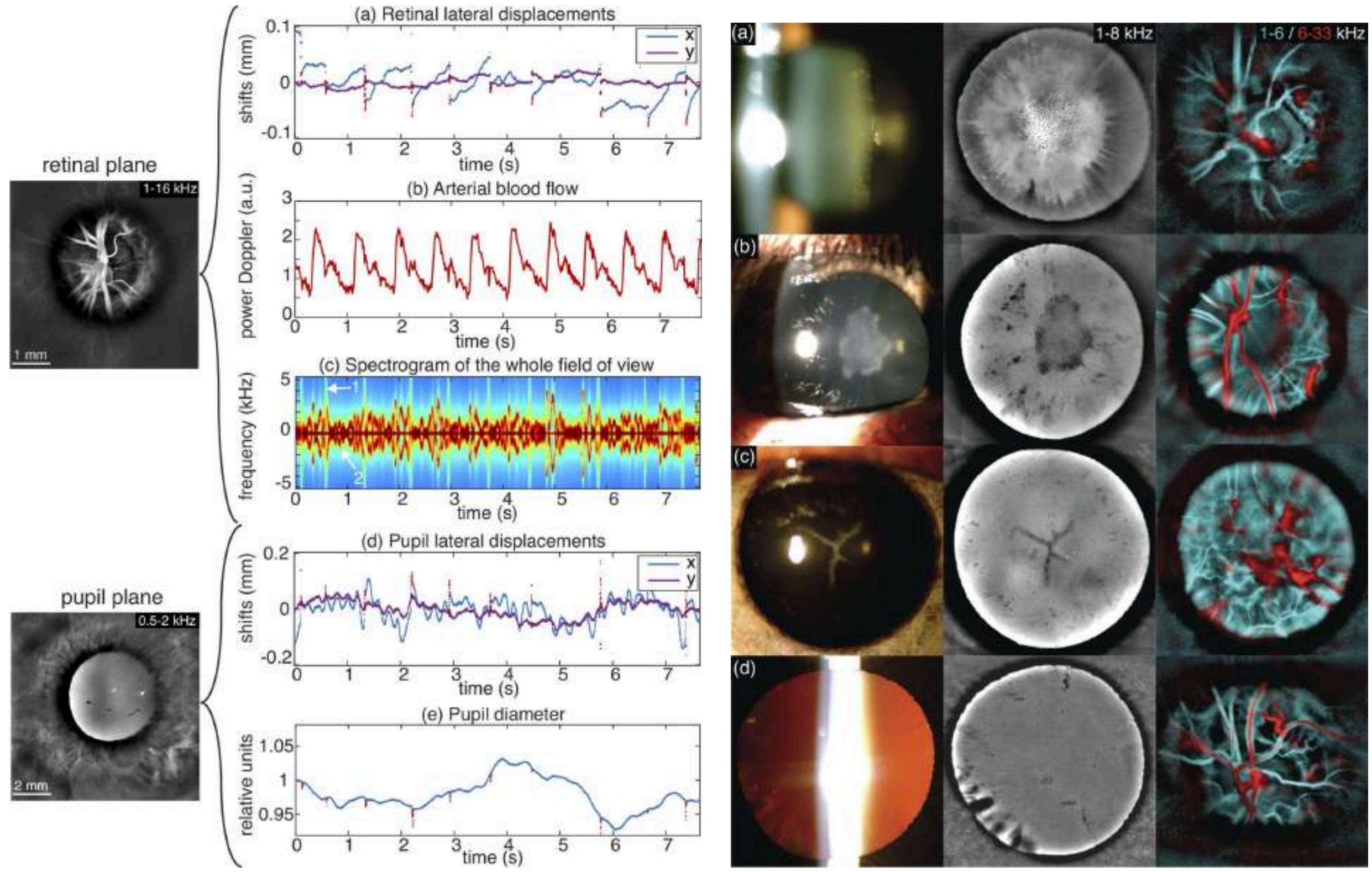

*Fig. 17 : **Left** : In the retinal plane are measured (a) the fundus lateral displacements, (b) the retinal arterial blood flow, and (c) the averaged Doppler spectrogram of the fundus revealing the signatures of in-plane saccades and out-of-plane motion. In the anterior segment are tracked (d) the pupil lateral displacements, and (e) the pupil diameter. **Right** : Imaging crystalline lens opacification. For each eye are shown the slit lamp examination (left), and from the same hologram data, the 1-8 kHz power Doppler image in the anterior segment (middle column), and the slow/fast flow color composite image in the posterior segment (right column). (a) Posterior subcapsular cataract. (b) Anterior polar cataract. (c) Star-shaped congenital cataract. (d) Presence of riders (early cataract).*

## Retinal blood flow reversal quantitatively monitored in out-of-plane vessels with laser Doppler holography

Doppler holography leverages the asymmetry of the Doppler spectrum to discern the local direction of blood flow in out-of-plane vessels, representing flow towards the camera in red and flow away in blue, mirroring the approach of ultrasound color

Doppler imaging [3]. The directional contrast provided by this technique enhances vessel topology understanding and aids in identifying tilted anatomy and sections of cilioretinal arteries entering the retina. It enables quantitative velocity monitoring through spectrograms based on spectral asymmetry in out-of-plane vessels (Fig. 18). Moreover, the technique allows for the unambiguous detection of pathological arterial flow reversals in diseased eyes, such as retrograde diastolic flows in glaucoma and vein occlusion cases. The color-encoded maps delineating blood flow direction furnish critical insights into vessel topology and flow and confidently reveal flow reversals that are not easily detectable using alternative optical techniques.

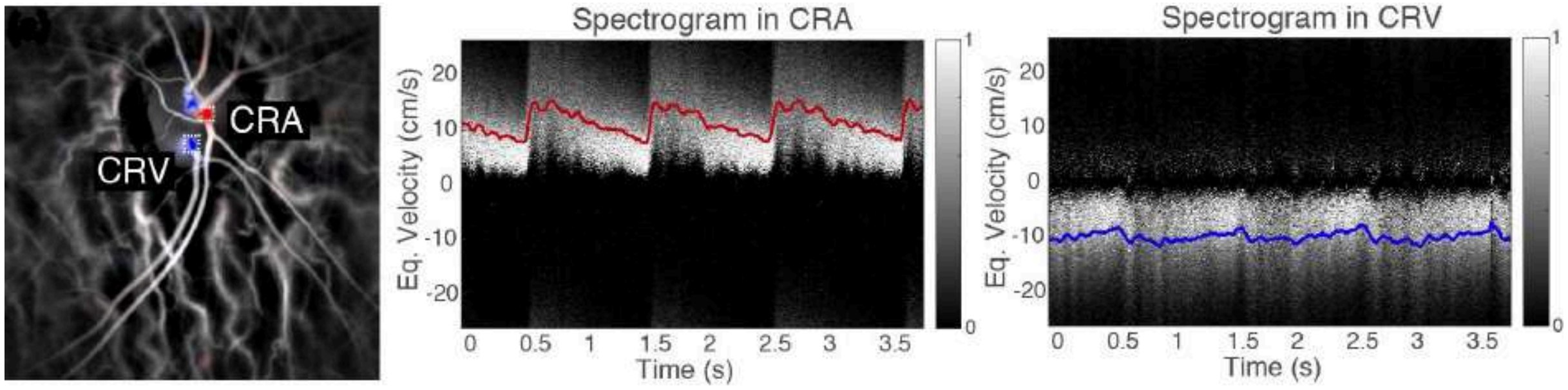

*Fig. 18 : The asymmetry of the Doppler optical fluctuation spectrum of complex-valued holograms is exploited to create directional color Doppler map (left) where red and blue reveal upward/downward flow . The velocity of the axial flow component in the central retinal artery and vein (CRA : center, CRV : right) can be estimated quantitatively in cm/s.*

## Diffuse laser illumination for Maxwellian view Doppler holography of the retina

Integrating diffuse illumination can significantly enhance the design of digital holographic imaging devices intended for ophthalmology [p2]. Introducing a diffusing element creates angular diversity in the optical radiation, dispersing the energy distribution of the illumination beam across the focal plane of the eyepiece (Fig. 19). The field of view for digitally computed retinal images can be readily expanded by positioning the eyepiece closer to the cornea. This adjustment facilitates a Maxwellian view of the retina without risking ocular harm. Adherence to American and European safety standards (ANSI Z136.1-2014 and ISO 15004-2 norms) for ophthalmic devices becomes more straightforward. It eliminates the issue of a laser hot spot in front of the cornea, typically observed without a scattering element, thus promoting safer wide-field retinal imaging via digital holography. Additionally, the utilization of diffuse laser illumination does not compromise the quality of digitally computed laser Doppler images.

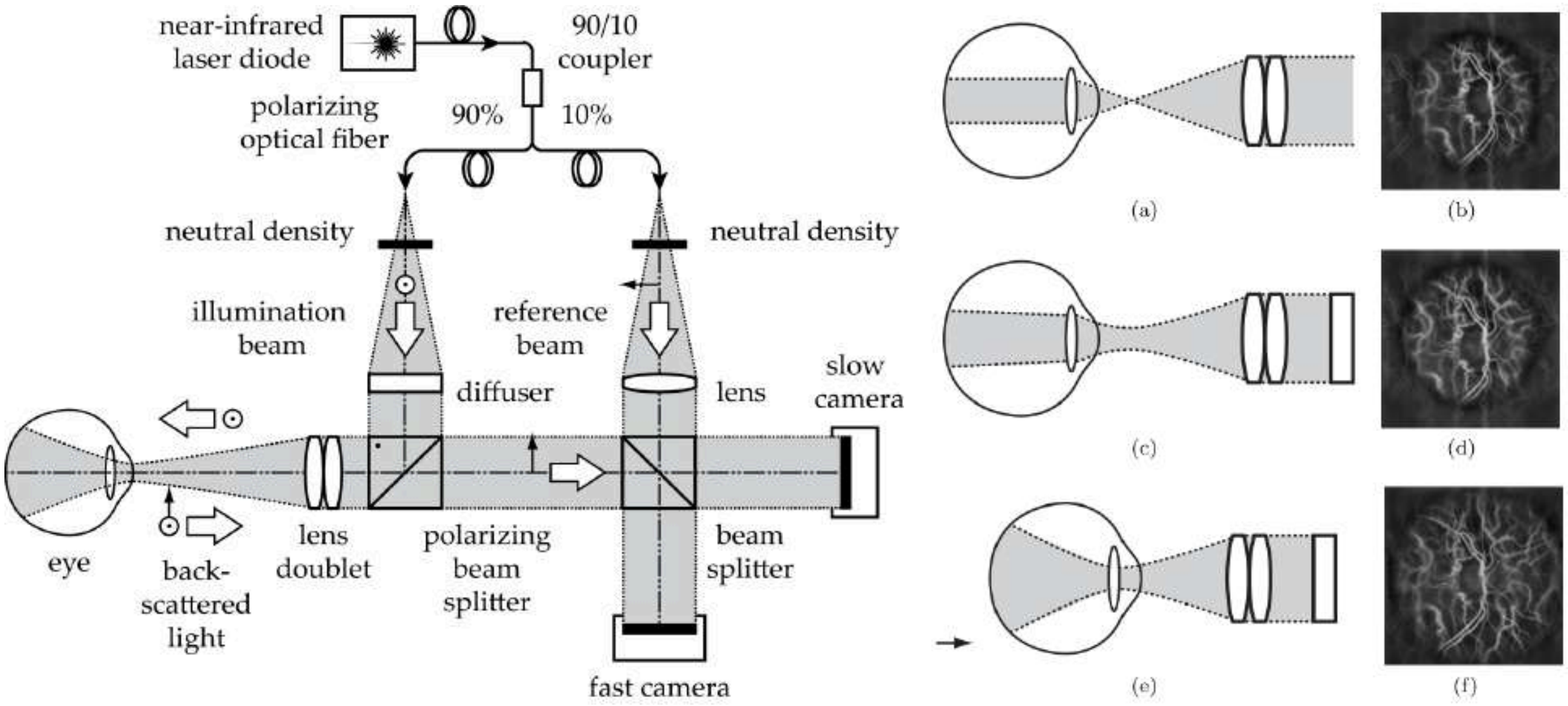

*Fig. 19* ***Left*** *: Doppler holography with diffuse laser illumination. An inline Mach-Zehnder near-infrared laser interferometer mixes the light backscattered by the eye fundus of a volunteer with a separate reference beam. Two cameras are set to record optical interference patterns. L1, L2 and L3 are converging lenses. The light source is a single wavelength laser diode. The 90% output of the fiber coupler is used for the object arm. The data from the fast camera is processed offline while the slow camera is used for real-time imaging. An optical diffuser scatters the illumination beam.* ***Right*** *: When the diffuser is not present (a), a hot spot (high local optical power density, narrow beam waist) is created at the laser focus in front of the cornea, and the iris acts as an image field diaphragm. When the diffuser is present (c), the input beam waist is large enough and spatially homogeneous to prevent any security hazard. This configuration offers the same field-of-view ((d) vs. (b)). The cornea of the patient's eye can be positioned in the focal region of the illumination beam (e), which increases the field of view (f) and the iris no longer acts as a field diaphragm.*

# Ongoing activities

## Rollout of digital holography devices in ophthalmology

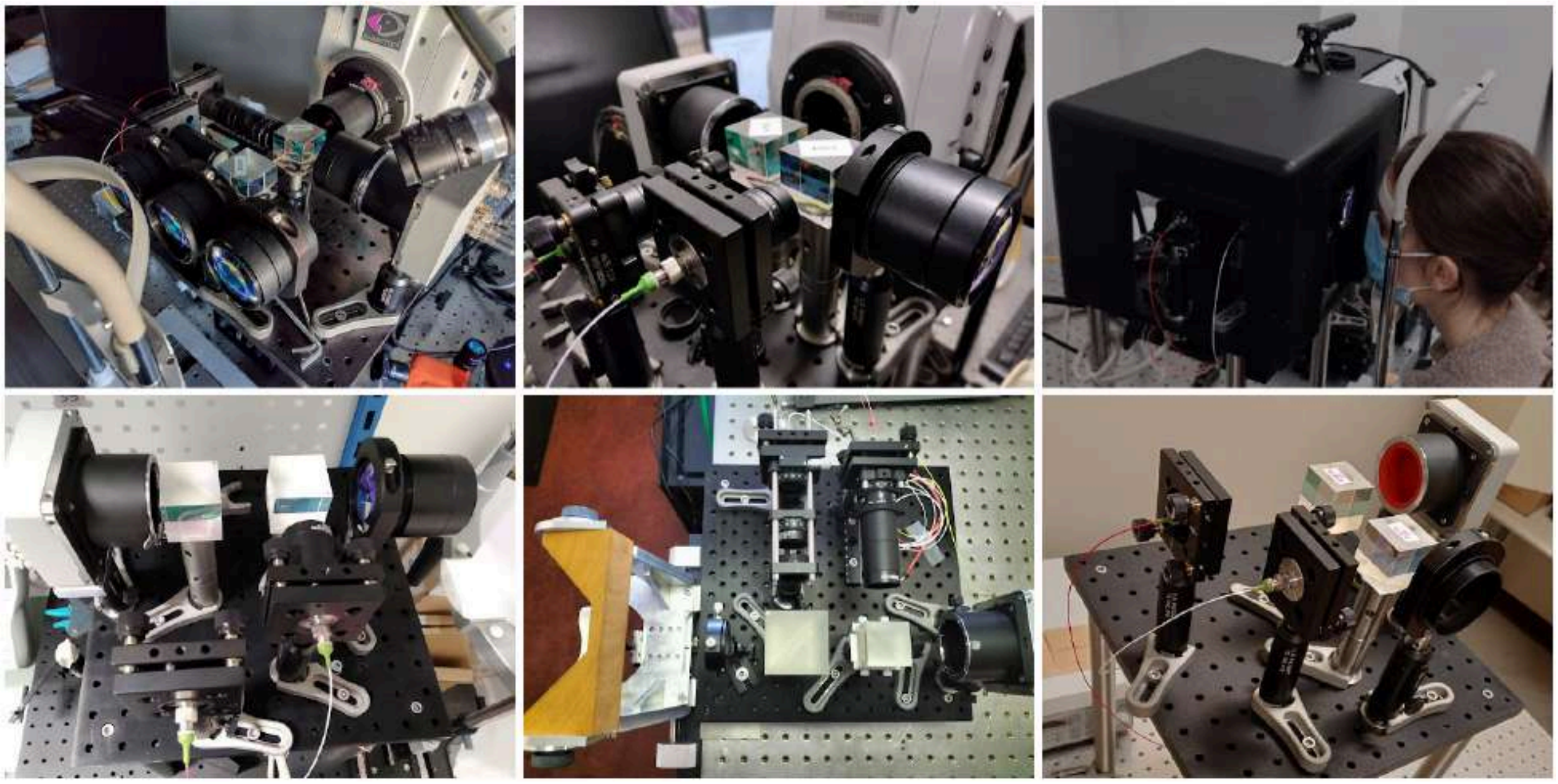

*Fig. 20 : Doppler holography prototype devices are now operational in various facilities: Quinze-Vingts (top left) and Rothschild Foundation (top center and top right) hospitals, Lumibird Quantel Medical (bottom left), ONERA Châtillon (bottom center), and the Advanced Ophthalmic Imaging Laboratory, UPMC, University of Pittsburgh (bottom right). The instrument's remarkable simplicity and robustness make it easy to replicate following publicly available conception guidelines and open source software.*

Prototypes of Doppler holography devices have been effectively integrated into a range of institutions, notably the Quinze-Vingts and Rothschild Foundation hospitals, as well as Quantel Medical, ONERA Châtillon, and the Advanced Ophthalmic Imaging Laboratory at UPMC, University of Pittsburgh (Fig. 20). To facilitate and foster these efforts, we are committed to offering continuous support through the provision of high-performance, regularly updated open-source software. In the latest iteration of the digital holography device designed for ophthalmology (as illustrated in Fig. 19), we have prioritized ease of implementation within clinical imaging settings. Notably, this instrument boasts remarkable stability and resilience against potential optical misadjustments over extended durations, ensuring consistent calibration. Furthermore, its capacity for highly repeatable Doppler measurements establishes a reliable foundation for clinical research endeavors. Doppler holography provides detailed imaging of choroidal vessels, rivaling indocyanine green angiography and offering complementary information. It can reveal and distinguish between deep choroidal arteries and veins (Figs. 12, 19, 21). Overall, Doppler holography provides non-invasive insights into the structure and function of the largest vessels in the retinal and choroidal vascular networks. Future performance improvements for Doppler imaging are expected to come primarily from upgraded software that integrates advanced signal processing, rather than from modifications to the already straightforward device.

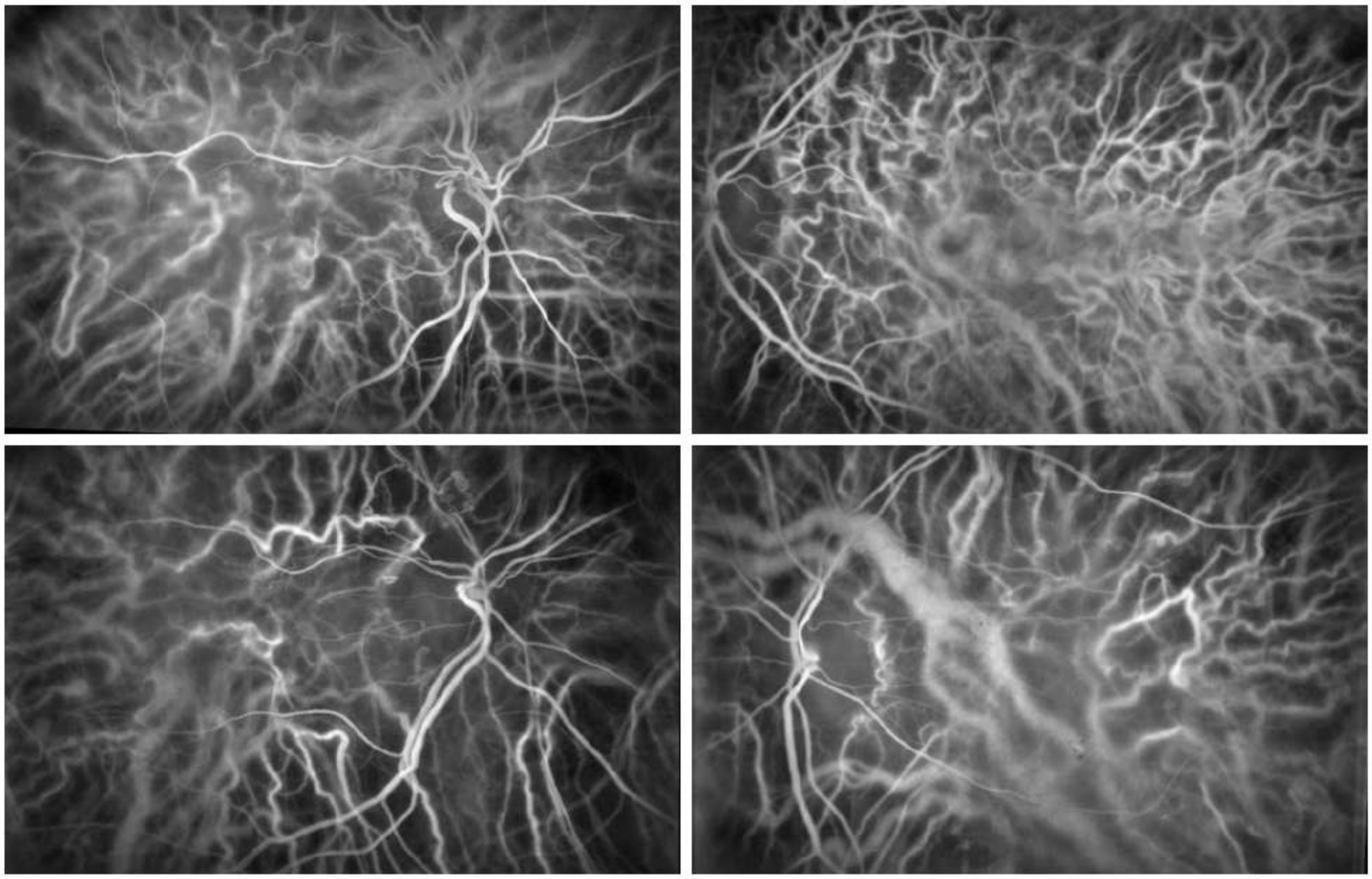

*Fig. 21 : Doppler images of the eye fundus of patients Doppler holography with diffuse laser illumination, revealing the main superficial vascular network of the retina and deeper choroidal vessels.*

The streamlined digital holography rendering software, Holowaves, along with its dependencies, will undergo further enhancements to maximize the usability of its source code. This software is designed to guarantee consistent quality and efficiency in Doppler image rendering and blood flow measurements (Fig.22) across various clinical research facilities. The reliability of Holowaves and its upcoming updates are paramount, as they will form the backbone for image rendering, signal processing, and data mining operations in clinical research centers.

The collection of Matlab scripts that make up Holowaves will be better organized to encourage the development of plugins by interns and PhD students specializing in physics, optics, and engineering. Initiated in 2018, this software project employs a wide range of dedicated functions tailored for each specific signal processing task. Recognizing the importance of clarity, we are planning comprehensive updates to our documentation to cater to prospective users and developers. The Holowaves software platform is open-sourced under a GPL3 license and is available on GitHub.

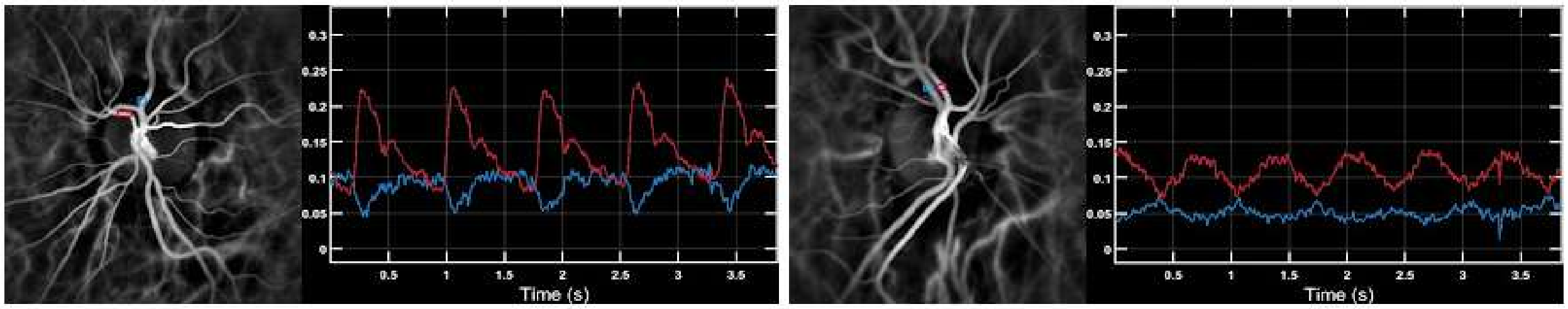

*Fig. 22 : Power Doppler signal (a.u.) in the retinal artery and vein of a control subject (left) and a patient with Takayasu's arteritis (vasculitis, right). Linked media shows the retinal pulse observed in these cases and in other pathologies: central retinal vein occlusion, arterial hypertension, glaucoma, and arrhythmia (movie).*

In high-resolution settings, Doppler image quality can be compromised by aberrations stemming from the eye's anterior segment and the eyepiece. To enhance image resolution, one can compensate for these aberrations by deconvolving the measured optical field using an estimated apparatus function. This strategy capitalizes on the interferometric measurement of quadrature-resolved optical wavefields. The aberration kernel's numerical estimation can be achieved using a digital Shack-Hartmann wavefront algorithm during the image rendering process, which approximately doubles the rendering time. The state-of-the-art subaperture correlation method's numerical phase correction algorithm has been successfully integrated into the Doppler image rendering routines of the holowaves software to facilitate this correction (Fig. 23).

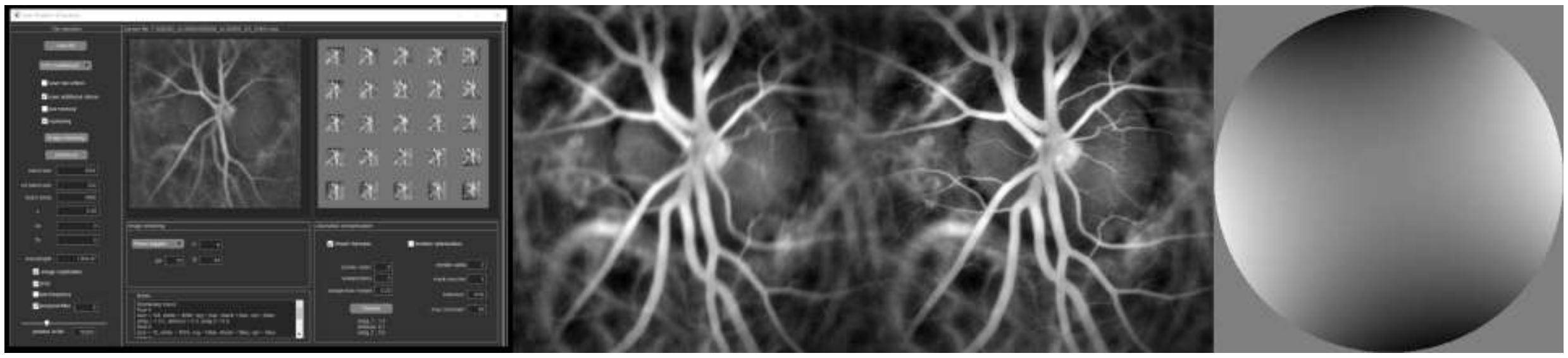

*Fig. 23 : Offline, fully automated Doppler image rendering (center) and aberration compensation from interferograms recorded at 67 kHz. Deterministic Shack-Hartmann wavefront analysis with the Holowaves open source software (left) enables automated digital rephasing (right) and reveals peripapillary vessels.*

## Real time, high throughput digital hologram rendering

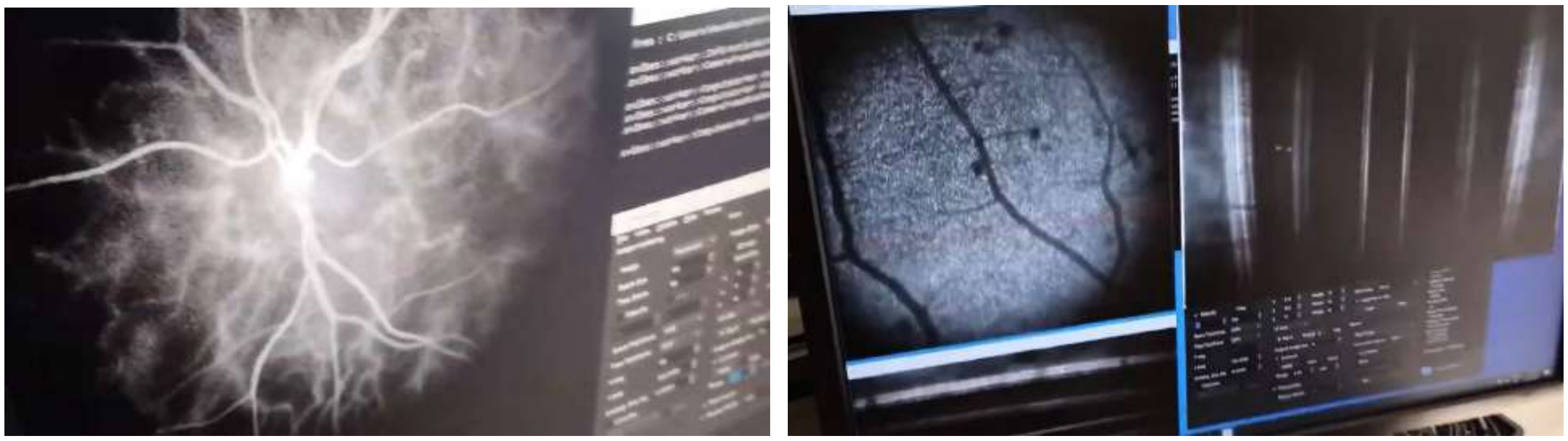

Fig. 24 : Real-time Doppler holography (movie) and holographic OCT (movie) at 20,000 frames per second (10 GB/s) with the software holovibes. OCT interferogram data is provided courtesy of Egidijus Auksorius and Maciej Wojtkowski.

In its current iteration, the versatile open source real-time hologram rendering software, holovibes, is capable of facilitating image rendering with the highest throughput cameras for real-time Doppler holography, holographic OCT, and other innovative imaging techniques (Fig. 24). To ensure its reliability and peak performance, the addition of new features will be constrained. The Holovibes software is engineered for real-time hologram image rendering in clinics involved in the Doppler holography project. It processes interferogram data from streaming cameras using parallel computing on Graphics Processing Units. At present, it can propagate the angular spectrum of 512-by-512 pixel interferograms at a remarkable rate of roughly 20,000 frames per second. This performance equates to a data throughput of about 10 GB/s, on par with the capabilities of the latest high bitrate streaming cameras. Seamless streaming is guaranteed without any frame drop, maintaining a maximum latency of approximately 30 ms. This advancement considerably improves patient examination processes.

This software will be updated to accommodate the forthcoming generation of high-bitrate streaming cameras and frame grabbers using transceivers based on CoaXPress (CXP) - a high-speed asymmetric communication protocol developed for seamless interfacing with video sensors via a single coaxial cable, widely used in industrial vision, and Quad Small Form-factor Pluggable communications standards (QSFP+, QSFP28, and QSFP56), used for the development of high-throughput streaming cameras. This endeavor will benefit from extensive collaborations with industrial partners, including Ametek and Euresys. Additionally, the software development of holovibes will continue to serve as an educational tool for students in the realm of high-performance computing.

## Digital holography of the anterior segment of the eye

Various techniques exist for imaging anterior segment blood vessels, but they all have their limitations. Currently, the most advanced method of angiography involves injecting fluorescent dyes, which is invasive and may be difficult to perform and quantify. OCT angiography has a long acquisition time, low resolution, and is unable to reveal blood flow fluctuations.

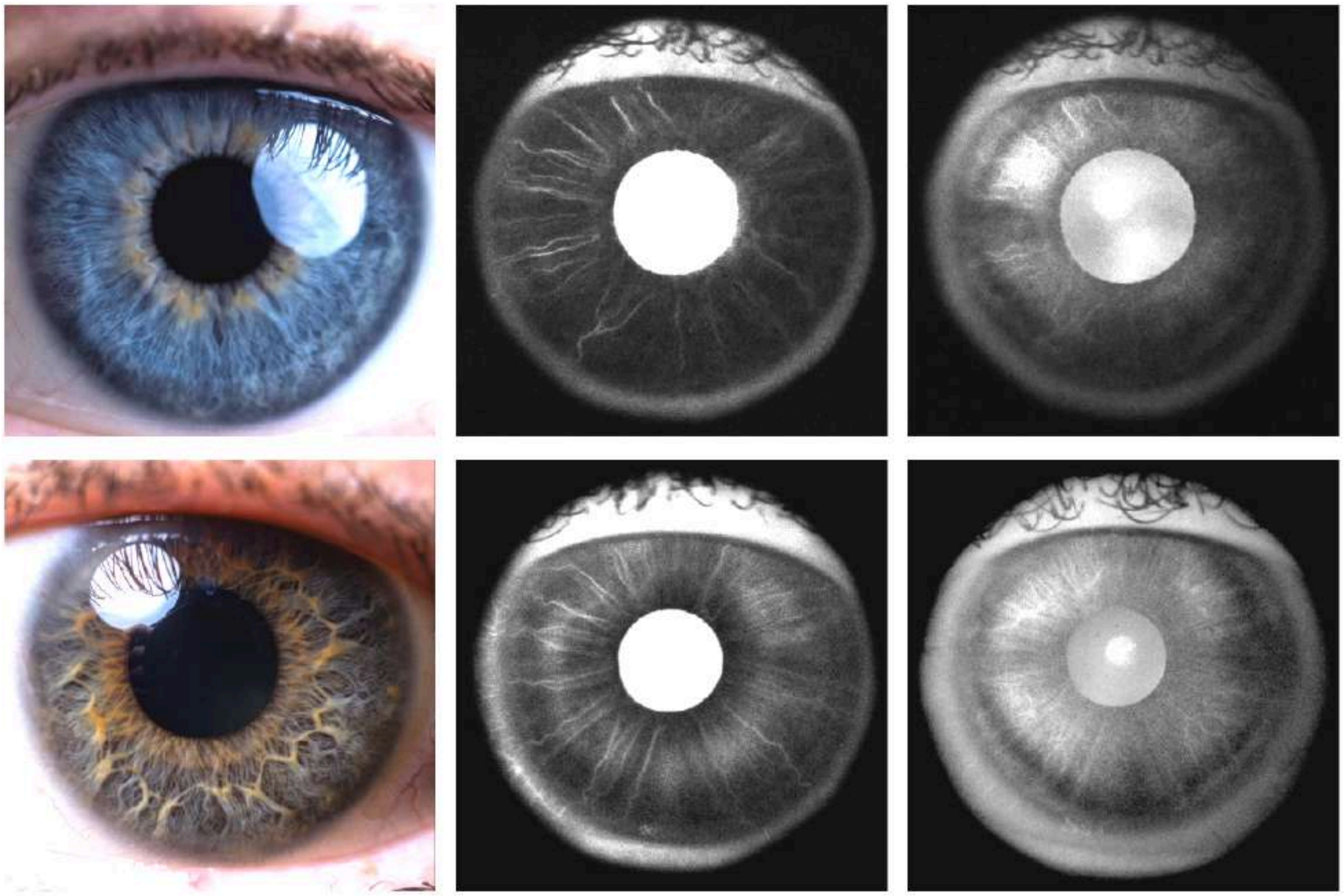

Fig. 25 : Photographs (left column) and real time Doppler images of the anterior segment of the eye. Doppler images with a high number of singular components reveal the anterior ciliary arteries (ACAs) forming the major arterial circle of the iris (right) and a low number of singular components reveal radial iris arteries and veins (center).

In addition to its ability to image retinal and choroidal blood flow, near-infrared Doppler holography can visualize in the superficial vasculature of the anterior segment of the eye in real time without any instrument modification. Pulsatile blood flow is visible in the  arteries of the conjunctiva, episclera, and iris in control patients with eye colors ranging from light blue to brown (Fig. 25). Several anatomical landmarks can be identified (Fig. 26), including ciliary arteries. Arterial pulse waves and venous flow were observed in vessels of various sizes, depending on the lateral field of view and the parameters of the principal components analysis. Important potential use cases arise because it is much less demanding than eye fundus imaging in terms of camera frame rate, probably because the sclera is much less vascularized than the choroid.

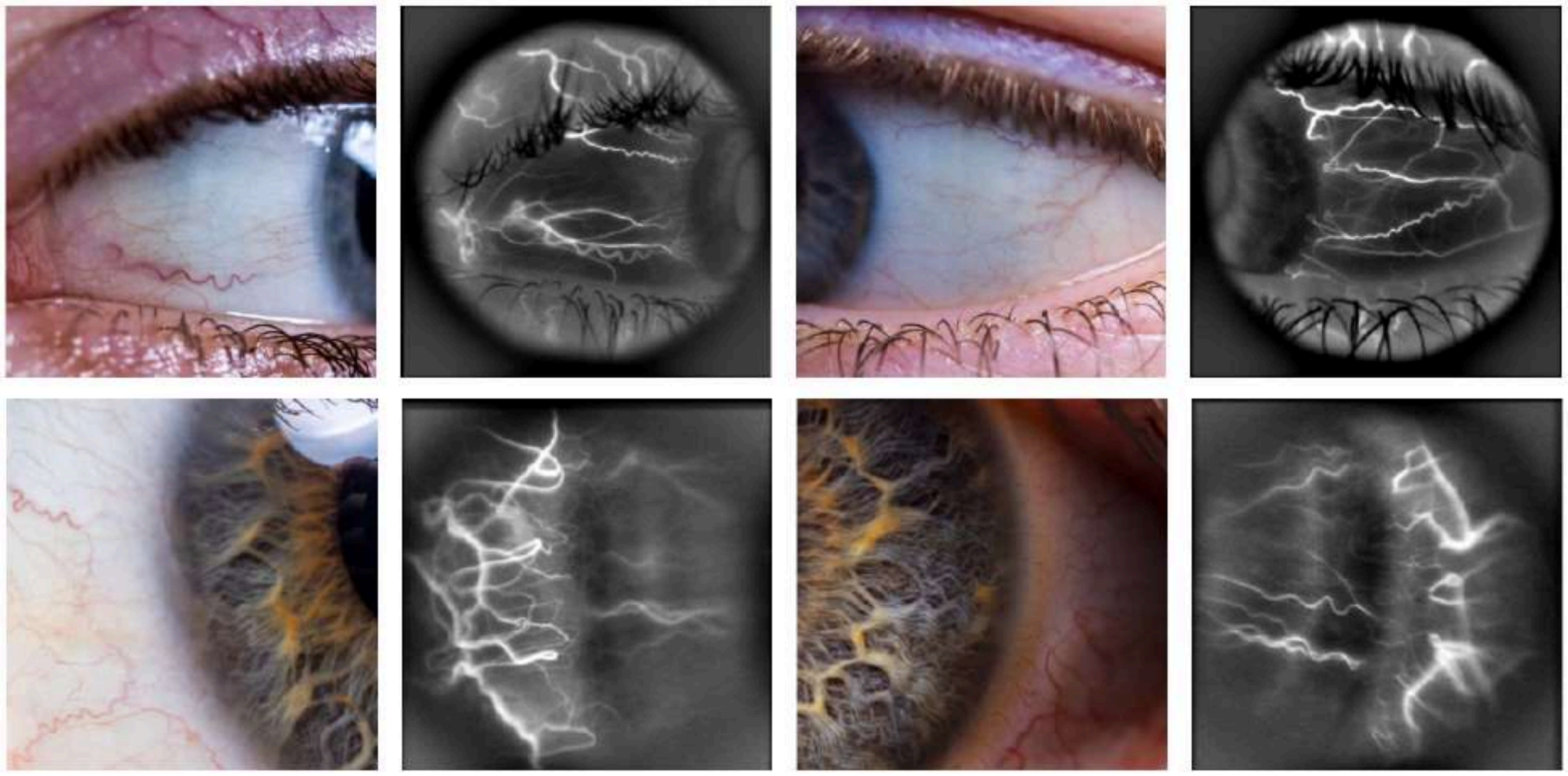

*Fig. 26 : Photographs and Doppler images of anterior conjunctival and episcleral arteries and veins along with anterior and long posterior ciliary arteries. Zoom on the corneal limbus in photographs revealing conjunctival and episcleral arteries. Doppler imaging reveals the transition between the sclera and the iris and some of the vessels that are going under the iris structure.*

A notable feature of the instrument is the ability to concurrently image both the anterior and posterior segments of the eye from the same set of recorded interferograms, as depicted in Figure 17. This capability suggests the potential for a versatile ophthalmic imaging device. The light that is backscattered from the retina retro-illuminates the anterior segment in transmission. This reveals any opacities in the cornea and eye lens that result from absorption or scattering. This arrangement offers an opportunity for investigating local phase contrasts within the anterior segment. The combination of phase-resolved optical field measurements by interferometry and optical channel transillumination of the anterior segment (Fig. 17) may facilitate the design of phase-contrast and coded-aperture imaging modalities in the future.

## Estimation of quantitative hemodynamical, rheological, and mechanical parameters from Doppler holograms

The research focus in Doppler holography is currently shifting towards the identification of novel functional biomarkers through numerical processing. These biomarkers are being designed to enhance non-invasive categorization of pathology severity and improve the effectiveness of therapeutic monitoring. Specifically, Doppler holography presents significant potential in evaluating and characterizing quantitative parameters, including blood velocity, blood volume rate, arterial resistivity index, blood viscosity, and arterial stiffness.

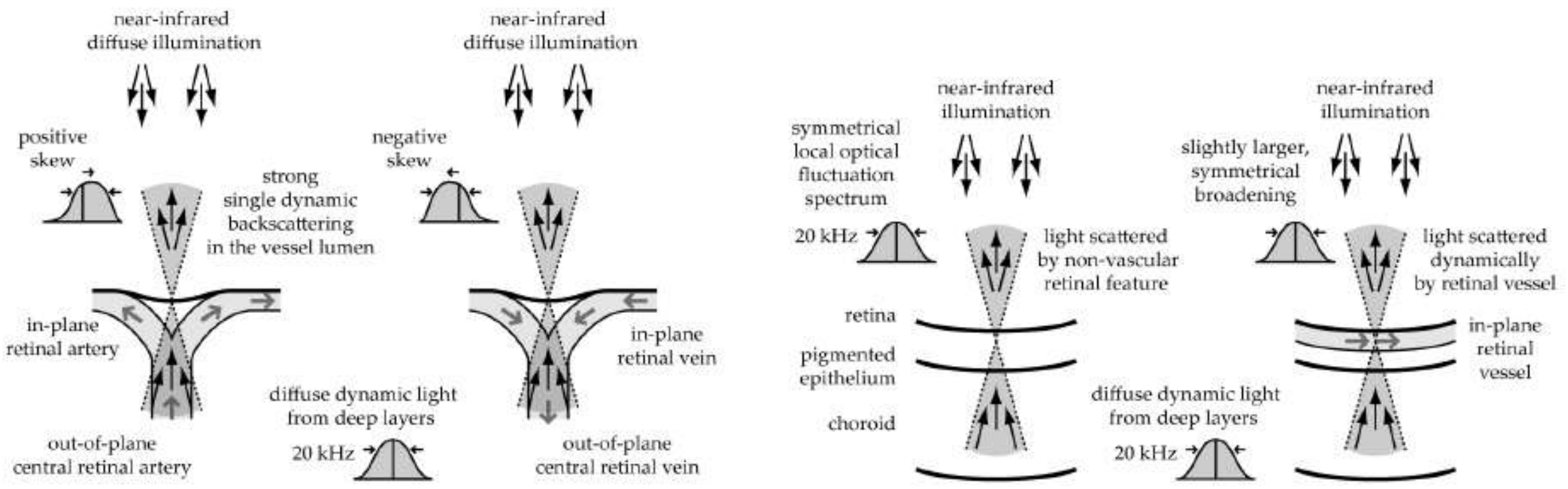

*Fig. 27 : A simple model of forward scattering of light from a diffuse, Doppler-broadened secondary source from the deep choroid can be used for parameterless estimation of blood flow in in-plane retinal vessels.*

A promising method for quantitatively assessing local blood velocity in superficial retinal vascular networks employs a straightforward model. This model is anchored in the forward scattering of light from a diffuse, Doppler-broadened secondary source that originates from the deep choroid. It facilitates a parameter-free estimation of blood flow in in-plane retinal vessels. The model operates under the premise that the optical field, as collected and measured via interferometry, emanates from a secondary light source that undergoes diffuse backscattering from both the sclera and the choroid. This results in a Doppler-broadened lineshape characterized by relatively smooth spatial variations. Blood flow in large superficial retinal vessels (ranging from 50 to 100 micrometers) introduces either asymmetrical or symmetrical additional broadening, as illustrated in Fig. 27.

The absolute estimation of blood flow in retinal vessels can be explored by measuring the differences between the width and shift of the Doppler lineshape in the vessels compared to the surrounding background. This approach is supported by a wealth of evidence from prior studies. Notably, this model presents as an efficient tool for gauging local blood flow velocity in prominent in-plane retinal vessels throughout the cardiac cycle. The local root-mean square velocity of blood flow can be deduced from the normalized second order moment of the frequency spectrum, the collection angle, and the locally accentuated Doppler broadening.

Consequently, this velocity estimation could serve as a means to evaluate blood volume rates within the retinal network by measuring vessel cross-sections. It can also assist in determining the arterial resistivity index by gauging the modulation depth of the velocity during the cardiac cycle, providing insights into the vascular network's resistance to blood flow. Given adequate spatial resolution, velocity profiles in large retinal vessels might pave the way for estimating the dynamic viscosity of blood. Such quantitative assessments will require meticulous design and validation, drawing from the expanding database of raw interferogram data gathered from both patients and control subjects.

# Selected references by the author

## Peer-reviewed journal articles

# Preprints

# Conference proceedings

## RÉSUMÉ

Une évaluation complète de la santé rétinienne nécessite des méthodes fiables et précises pour mesurer la perfusion sanguine localisée. Malgré des avancées considérables dans les techniques d'imagerie telles que l'angiographie au vert d'indocyanine et à la fluorescéine, ainsi que l'angiographie par tomographie en cohérence optique, leur capacité à surveiller la dynamique du flux sanguin tout au long du cycle cardiaque présente d'importantes limitations. Pour une prise en charge plus efficace des personnes souffrant de conditions oculaires, innover en proposant de nouvelles approches est primordial. L'holographie Doppler, une technique d'imagerie optique non invasive émergente, relève ce défi en offrant une imagerie à haute résolution temporelle du flux sanguin rétinien et choroïdien. Désormais un domaine de recherche interdisciplinaire en plein essor, l'holographie Doppler entrelace la conception de systèmes d'imagerie optique fonctionnelle, l'informatique haute performance et l'investigation clinique. Grâce aux efforts collaboratifs entre universités, partenaires industriels et cliniques ophtalmologiques, un réseau pour son développement et son application se forme. Cette initiative promet de stimuler la découverte de nouveaux biomarqueurs fonctionnels, transformant le diagnostic et le traitement des maladies rétiniennes, affinant la catégorisation de la gravité des maladies, et améliorant la surveillance thérapeutique - conduisant finalement à de meilleurs résultats de soins de santé.

## ABSTRACT

A comprehensive assessment of retinal health demands reliable and precise methods to measure localized blood perfusion. Despite considerable advancements in imaging techniques, such as indocyanine green and fluorescein angiography, along with optical coherence tomography angiography, their capacity to monitor blood flow dynamics across the cardiac cycle faces significant limitations. For more effective care of those with ocular conditions, innovating new approaches is paramount. Doppler holography, an emerging non-invasive optical imaging technique, meets this challenge by offering high temporal resolution imaging of retinal and choroidal blood flow. Now a burgeoning interdisciplinary research field, Doppler holography intertwines functional optical imaging system design, high-performance computing, and clinical investigation. Through collaborative efforts among universities, industry partners, and ophthalmic clinics, a network for its advancement and application is taking shape. This endeavor promises to propel the discovery of novel functional biomarkers, transforming the diagnosis and treatment of retinal diseases, refining disease severity categorization, and enhancing therapeutic monitoring—ultimately leading to improved healthcare outcomes.